\documentclass[reprint,amsmath,amssymb,aps,prb,superscriptaddress,floatfix,nofootinbib]{revtex4-2}

\pdfoutput=1
\usepackage{graphicx}
\usepackage{dcolumn}
\usepackage{bm}
\usepackage{braket}
\usepackage[mathlines]{lineno}
\usepackage{stackengine}
\usepackage{verbatim}
\usepackage{graphics}
\usepackage{hyperref}
\usepackage{xcolor}
\usepackage{appendix}
\usepackage{soul}
\usepackage{ulem} 

\begin{document}
\preprint{APS/123-QED} \title{Interplay of many-body interactions and
  quasiperiodic disorder in the all-band-flat diamond chain}

\author{Aamna Ahmed}
\affiliation{Department of Physics, Indian Institute of Science Education and Research, Bhopal, Madhya Pradesh 462066, India\\}
\author{{Nilanjan Roy}}%
\affiliation{Department of Physics, Indian Institute of Science, Bangalore 560012, India\\}
\author{Auditya Sharma}%
\affiliation{Department of Physics, Indian Institute of Science Education and Research, Bhopal, Madhya Pradesh 462066, India\\}

\date{\today}


\begin{abstract}
We study the effects of quasiperiodic Aubry-Andr\'e (AA) disorder and
interactions on a one-dimensional all-band-flat (ABF) diamond
chain. We consider the application of disorder in two ways: a
symmetric one, where the same disorder is applied to the top and
bottom sites of a unit cell, and an antisymmetric one, where the
disorder applied to the top and bottom sites are of equal magnitude
but with opposite signs. The single-particle wave-packet dynamics for
the clean system and when the disorder is applied symmetrically show
quantum caging; in the antisymmetric case, the wave-packet spreads
over the entire lattice. These results agree with our previous work,
where compact localization was observed in the case of the clean
system and for symmetrically disordered diamond lattices. In the
presence of nearest-neighbour interactions, nonergodic phases are
observed in the case of a clean system and symmetrical disorder; at
higher disorder strengths, we find an MBL-like phase in the symmetric
case. However, many-body non-equilibrium dynamics of the system from
carefully engineered initial states exhibit quantum caging. In the
antisymmetric case, a nonergodic mixed phase, a thermal phase and an
MBL-like phases, respectively, are observed at low, intermediate and high
disorder strengths. We observe an absence of caging and initial state
dependence (except at the intermediate disorder strength) in the study of non-equilibrium dynamics.
\end{abstract}

\maketitle


\section{INTRODUCTION}

Flat band (FB) systems, which are characterized by highly degenerate
energy levels and support \textit{compact localized eigenstates}
(CLS)~\cite{intro4, Sathe_2021}, has been a subject of great interest
over the last decade~\cite{PARAMESWARAN2013816,
  doi:10.1142/S021797921330017X, intro1, intro2, intro3}, although the
concepts are older where the term Aharanov Bohm (AB)
caging~\cite{PhysRevLett.85.3906, PhysRevLett.88.227005,dice1,dice2}
has been used. Compact localized states span strictly over a few unit
cells, with zero probability amplitude elsewhere in contrast to
Anderson localization~\cite{RevModPhys.80.1355}, where the `spread' of
a state dies down exponentially. While Anderson localization observed
in non-interacting disordered systems is now a mature topic with a
large body of literature around it, the localization characteristics
of quantum systems in the presence of both disorder and
interactions~\cite{PhysRevB.21.2366, PhysRevLett.78.2803,
  PhysRevLett.79.1837, BASKO20061126, RevModPhys.83.863, PhysRevB.104.155137} is an
actively evolving area of research. A prominent example is the
phenomenon of \textit{many-body localization} (MBL)~\cite{
  RevModPhys.91.021001,
  PhysRevB.82.174411,doi:10.1146/annurev-conmatphys-031214-014726,
  PhysRevB.87.134202,PhysRevB.91.184202, PhysRevB.92.100305,
  PhysRevB.96.035153} where the system fails to thermalize even in the
presence of interactions. Translationally invariant single-particle
flat band networks coupled with many-body interactions have also
recently gained a lot of
attention~\cite{TillekeDaumannDahm+2020+393+402, PhysRevB.102.041116,
  PhysRevB.103.L060301, PhysRevB.104.085132, PhysRevB.104.144207,
  PhysRevB.105.L041113}. These models exhibit nonergodic behaviour
with a lack of transport of particles for any interaction strength
exhibiting \textit{many-body flat band localization}
(MBFBL)~\cite{PhysRevB.102.041116, PhysRevB.105.L041113}. This
naturally motivates the study of flat band systems subjected to both
disorder and interactions~\cite{PhysRevResearch.2.043395,
  PhysRevB.105.094201, PhysRevB.104.094202}.

In one of our previous works~\cite{PhysRevResearch.2.043395}, we
systematically investigated the effects of turning on interactions in
the presence of uniform disorder on the all-band-flat (ABF) diamond
chain. This model shows a nonergodic mixed phase at low disorder
strength, separated from the MBL phase at high disorder strength by a
thermal phase at intermediate disorder strength. The addition of
disorder to flat-band systems is known to yield exotic
behavior~\cite{PhysRevLett.96.126401, PhysRevB.98.184202,
  PhysRevB.106.205119}. In our recent work~\cite{PhysRevB.106.205119},
we investigated the effect of a quasiperiodic Aubry-Andr\'e ($AA$)
on-site disorder~\cite{AA, Modugno_2009} on the ABF diamond chain. We 
found that the symmetry of the applied external potential plays a
crucial role. With a symmetric disorder, it is possible to completely
destroy the degeneracy and still preserve the compact localization of
the eigenstates~\cite{fbl2}. However, when the disorder is applied in
an antisymmetric fashion, both the degeneracy and compact localization
are destroyed and a robust \textit{flat-band-based
  multifractality}(FBM)~\cite{Kravtsov_2015, PhysRevE.98.032139,
  Facoetti_2016} is observed in an extensive
region of the phase diagram. In the present work, we study the effects
of interactions on the ABF diamond chain both in the absence and
presence of quasiperiodic disorder.

We begin by exploring single-particle dynamics, which shows quantum
caging in the long time limit for both the zero-disorder and symmetric
disorder cases. However, when the disorder is applied in an
antisymmetric manner since the compact localization of the
eigenstates is destroyed~\cite{PhysRevB.106.205119}, the time-evolved
state also displays a spreading over all the lattice sites. We next
investigate the properties of the clean system when interactions are
turned on. The system manifests nonergodic
phases at all interaction strengths in the zero-disorder case. However, from a study of
non-equilibrium dynamics, we conclude that for some specially
engineered initial states, many-body systems exhibit caging behaviour
independent of the strength of the interaction.

In the simultaneous presence of disorder and interactions, the
symmetry of the applied disorder is again crucial. A symmetric
disorder coupled with interactions yields nonergodic phases in the low
and intermediate disorder regimes and MBL-like behaviour in the high
disorder regime. We find that the dynamics is dependent on the initial
state; in particular, we observe quantum caging for specific
engineered initial states. The antisymmetric application of disorder
leads to a mixed nonergodic phase at low disorder strength, a thermal
phase at intermediate disorder strength, and an MBL-like phase at high
disorder strength. The mixed phase obtained at low disorder strength
is attributed to the presence of multifractal states in the single
particle limit. Although we find initial state dependence in the
non-equilibrium dynamics (except for intermediate disorder strengths, which yield a thermal phase), no quantum caging behaviour is seen.

This paper is organized as follows. In Section~\ref{sec:level2}, we
discuss the details of the model. In Section~\ref{sec:level3}, we
discuss the effects of $AA$ disorder on the single-particle dynamics
in the disorder-free, symmetric and antisymmetric cases.
Section~\ref{sec:level4} discusses the effects of interactions on the
clean ABF diamond chain. Section~\ref{sec:level5} explores the
symmetric application of quasiperiodic $AA$ disorder
on the interacting system. Section~\ref{sec:level6} discusses the
interplay of antisymmetric application of disorder 
and interactions. We then summarize our results in
Section~\ref{sec:level7}.


\section{Model}\label{sec:level2}
%
\begin{figure}
\centering
\stackunder{}{\includegraphics[width=8cm]{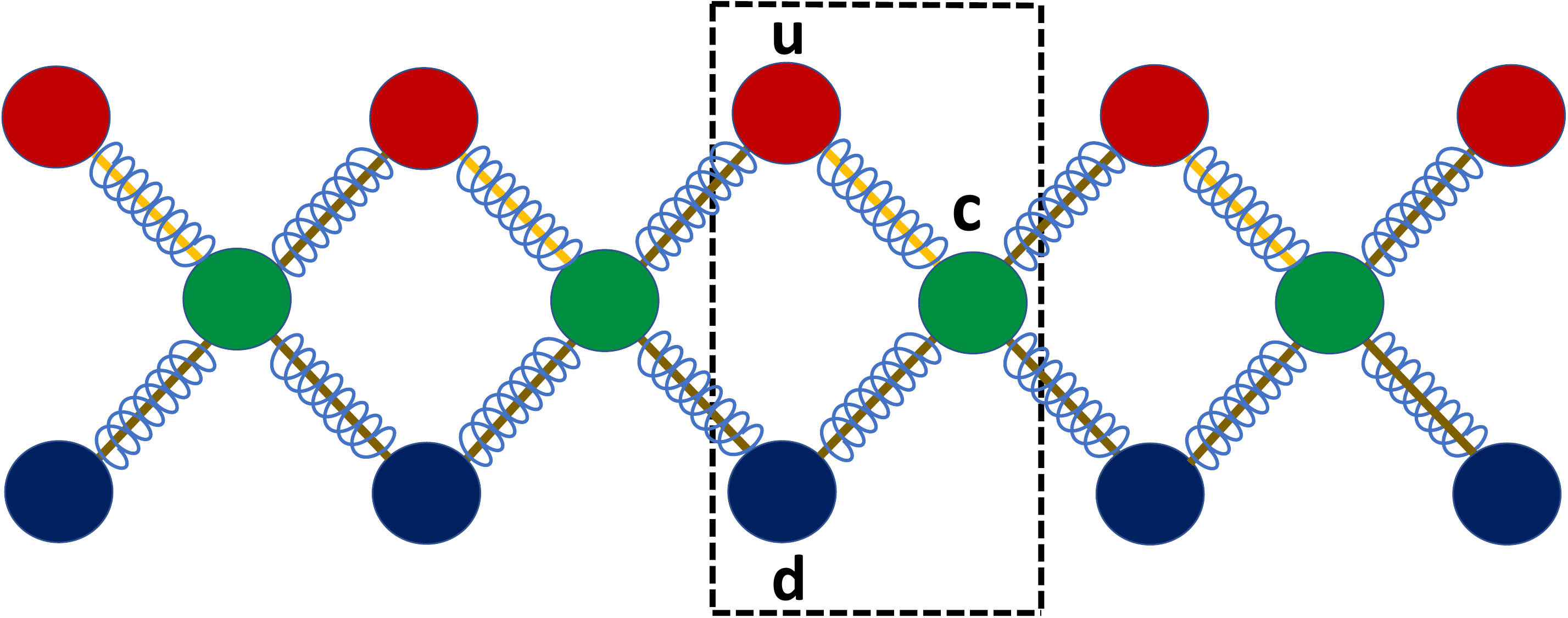}}
\caption{\label{fig1} Schematic representation of the diamond lattice with the $u$ (up), $d$ (down) and $c$ (centre) sites of a representative unit cell confined by the black dashed lines. Nearest neighbour interaction $V$ is represented by wiggly blue lines.}
\end{figure}


We study the ABF diamond lattice, where the $k^{\text{th}}$ unit cell
consists of three sites $\alpha_k=\left\lbrace u_k, d_k,
c_k\right\rbrace$ (see Fig.~\ref{fig1}).  The fermionic creation
operators acting at the $u$ (up), $c$ (center), and $d$ (down) sites
respectively in the $k^{\text{th}}$ unit cell are
$\hat{u}_{k}^{\dagger}, \hat{c}_{k}^{\dagger}$, and
$\hat{d}_{k}^{\dagger}$ and the Hamiltonian is:
\begin{equation}
 \hat{H}= \hat{H}_{\text{hop}}+ \hat{H}_{\text{os}}+ \hat{H}_{\text{int}},
\end{equation}
where
\begin{align}
\nonumber \hat{H}_{\text{hop}}=&-J \sum_{k=1}^{N / 3}\left(-\hat{u}_{k}^{\dagger} \hat{c}_{k}+\hat{d}_{k}^{\dagger} \hat{c}_{k}+\hat{c}_{k}^{\dagger} \hat{u}_{k+1}+\hat{c}_{k}^{\dagger} \hat{d}_{k+1}+\text{H.c.}\right)\\
\nonumber \hat{H}_{\text{os}}=& \sum_{k=1}^{N / 3}\left(\zeta_{k}^{u} \hat{u}_{k}^{\dagger} \hat{u}_{k}+\zeta_{k}^{c} \hat{c}_{k}^{\dagger} \hat{c}_{k}+\zeta_{k}^{d} \hat{d}_{k}^{\dagger} \hat{d}_{k}\right)\\
\nonumber \hat{H}_{\text{int}}=& V \sum_{k=1}^{N / 3}\left( \hat{u}_{k}^{\dagger} \hat{u}_{k} \hat{c}_{k}^{\dagger} \hat{c}_{k}+\hat{d}_{k}^{\dagger}\hat{d}_{k} \hat{c}_{k}^{\dagger}\hat{c}_{k}+\hat{c}_{k}^{\dagger} \hat{c}_{k}\hat{u}_{k+1}^{\dagger}\hat{u}_{k+1} \right.\\
& +\left. \hat{c}_{k}^{\dagger}\hat{c}_{k}\hat{d}_{k+1}^{\dagger} \hat{d}_{k+1}\right).
\label{eq1}
\end{align}
The total number of lattice sites is denoted by $N$, which should be a
multiple of $3$ owing to the unit cell structure of the periodic
lattice. The hopping amplitude is $J$, which is taken to be $1$ for
simplicity, and $V$ is the strength of the nearest neighbour
interaction. For each site of the $k^{\text{th}}$ unit cell, we
include independent on-site Aubry-Andr\'e potentials
\begin{equation}
\zeta_k^\alpha=\lambda_{\alpha}\cos(2\pi kb+\theta_p),
\label{eq2}
\end{equation}
where the strength of the potential is $\lambda_{\alpha}$ and the
quasi-periodicity parameter $b$ is taken to be the golden mean
$(\sqrt{5}-1)/2$. The arbitrary global phase $\theta_p$ is chosen
randomly from a uniform random distribution $[0,2\pi]$. Here we
consider two types of correlations between the on-site energies on the
up `$u$' and down `$d$' sites: a symmetric configuration in which
$\zeta_{k}^{u}=\zeta_{k}^{d}$ and an antisymmetric configuration in
which $\zeta_{k}^{u}=-\zeta_{k}^{d}$.

\begin{figure}[b]
\centering
\stackunder{\hspace{-4cm}(a)}{\includegraphics[width=4.3cm]{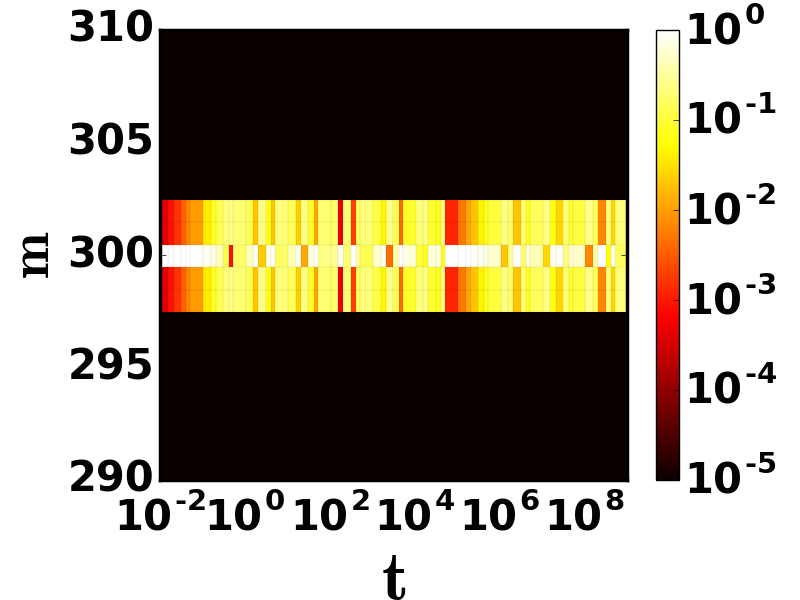}}\hspace{-1mm}
\stackunder{\hspace{-4cm}(b)}{\includegraphics[width=4.2cm]{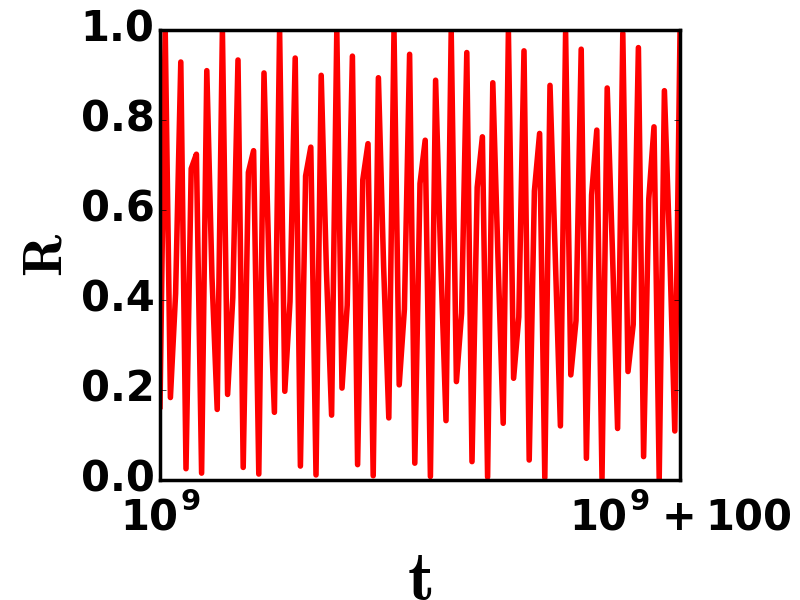}}
\vspace{-0.7cm}

\stackunder{\hspace{-4.2cm}(c)}{\includegraphics[width=4.3cm]{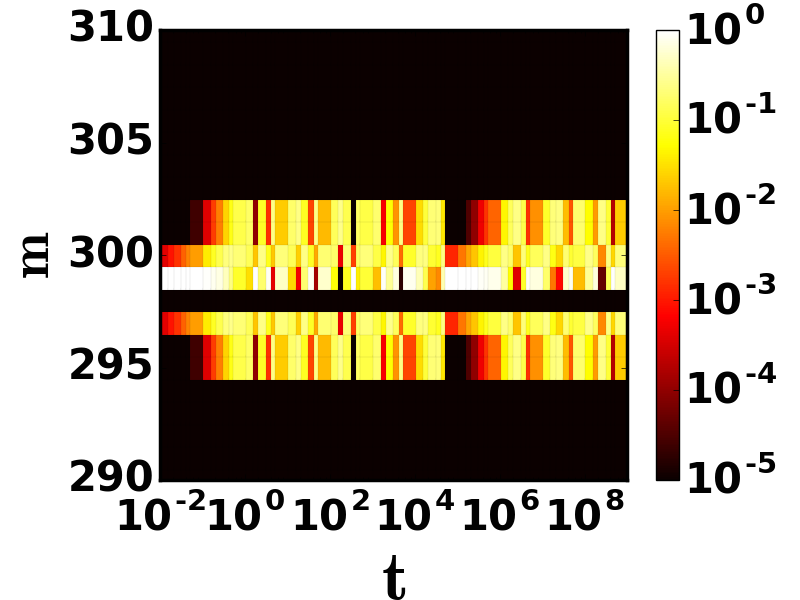}}\hspace{-1mm}
\stackunder{\hspace{-4.1cm}(d)}{\includegraphics[width=4.2cm]{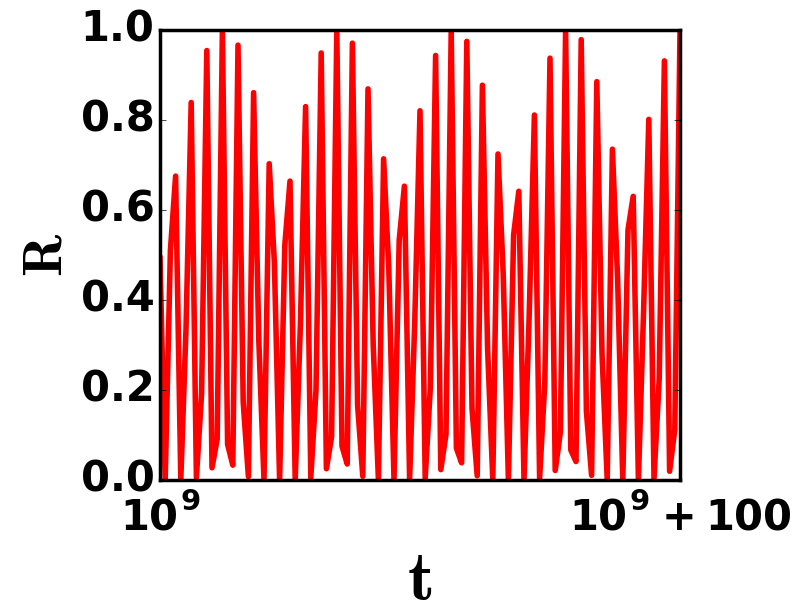}}
\caption{\label{fig2} Clean system: The particle density (whose value
  is represented by a colour according to the code shown) as a
  function of time $t$, with $m$ denoting the site index, for a
  single-particle initially at the (a) $c-$site and (c) $d-$site of
  the $100^{\text{th}}$ unit cell for system size $N = 600$ ($200$
  unit cells) in the ABF diamond lattice. The return probability
  $R(t)$ as a function of time $t$ for a particle initially at the (b)
  $c-$site and (d) $d-$site of the $100^{\text{th}}$ unit cell.}
\end{figure}

In the clean non-interacting limit, the ABF diamond chain possesses
three flat bands at energies $\pm 2, 0$ and no dispersive
band. Consequently, the system is a good insulator, possessing only
compact localized eigenstates. The system is highly degenerate, with
the CLS occupying two unit cells. The other states corresponding to
each flat band can be obtained by translating by an integer multiple
of unit cells along the lattice. In the presence of symmetric
disorder, remarkably, the eigenstates continue to be compactly
localized in the original basis~\cite{PhysRevB.106.205119} although
the translation symmetry and, thus, the flat band structure are broken.
On the other hand, when the potential is applied in an antisymmetric
manner, we find neither degeneracy nor compact
localization~\cite{PhysRevB.106.205119}, but a novel kind of {\it
  flat-band-based multifractality}.

\begin{figure*}
\centering
\stackunder{}{\includegraphics[width=0.75\textwidth]{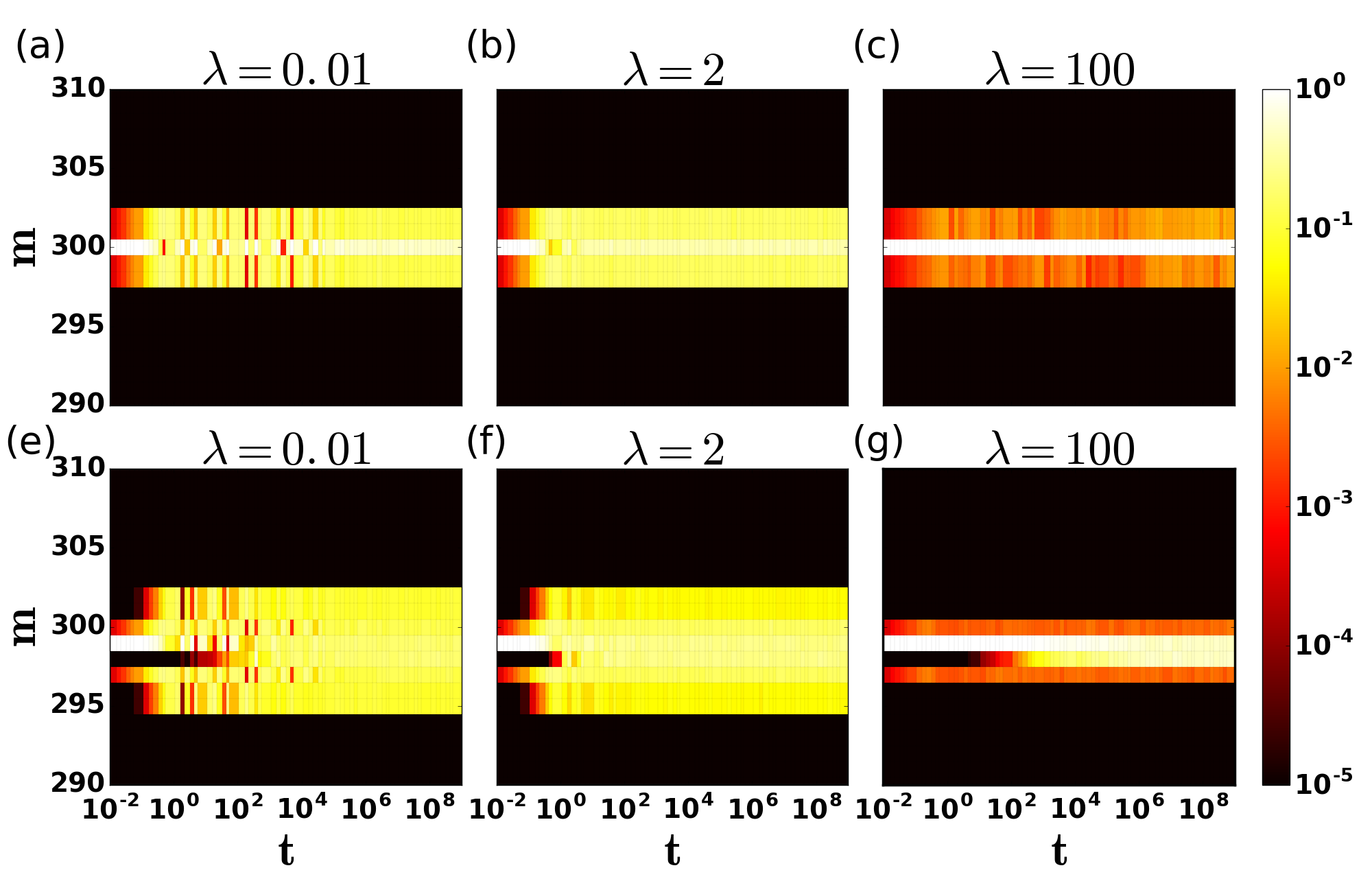}}\hspace{-3mm}
\stackunder{}{\includegraphics[width=0.25\textwidth]{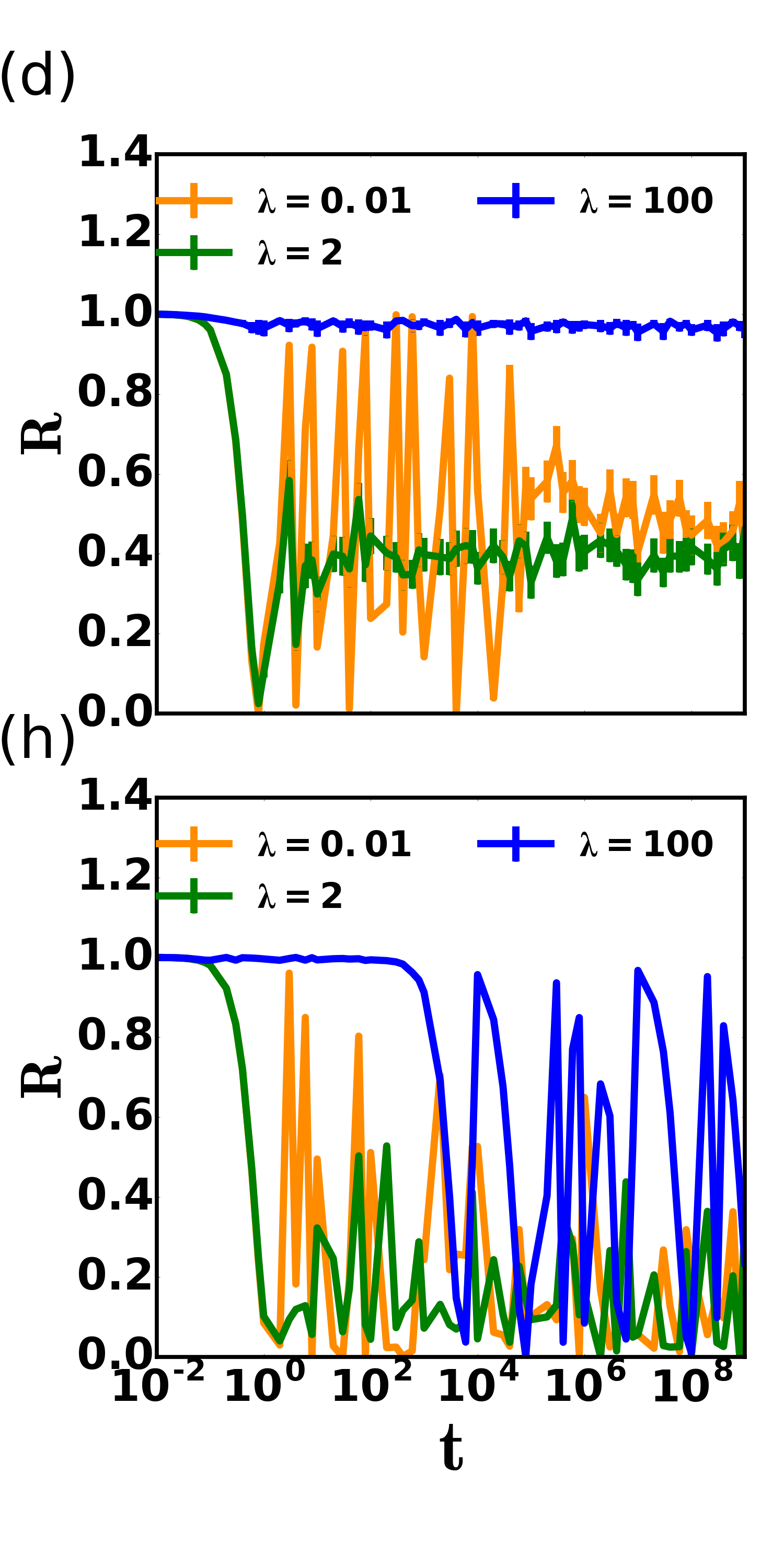}}
\caption{\label{fig3} In the symmetric case, the particle density (whose value is represented by a colour according to the code shown) as a function of time $t$, with $m$ denoting the site index, for a single particle initially at the $c-$site of the $100^{\text{th}}$ unit cell with increasing disorder strength (a)~$\lambda=0.01$, (b)~$\lambda=2$ and (c)~$\lambda=100$ and (d)~evolution of the return probability $R$. For a single particle initially at the $d-$site of the $100^{\text{th}}$ unit cell, the particle density as a function of time $t$ for (e)~$\lambda=0.01$, (f)~$\lambda=2$ and (g)~$\lambda=100$ and (h)~evolution of the return probability $R$. Here system size is $N = 600$, and the number of disorder realizations is $100$.}
\end{figure*}

\section{SINGLE-PARTICLE DYNAMICS}\label{sec:level3}
In this section, we explore the single-particle properties with the
help of non-equilibrium dynamics of the particle density
and the return probability. We will see that these results are
consistent with the static properties of the eigenstates we obtained
in our earlier study~\cite{PhysRevB.106.205119}. We study the dynamics
by considering two initial states, one where the single-particle
occupies the lattice site `$c$' of the $k^{th}$ unit cell
($\ket{\psi_{in}}=\ket{c_k}$) and the other where it occupies the
lattice site `$d$' of the $k^{th}$ unit cell
($\ket{\psi_{in}}=\ket{d_k}$).  We choose $k=N/6$ so as to focus on
the sites of the central unit cell - the total number of unit cells is
$N/3$. Once the initial state is fixed, we obtain the time evolved
state at time $t$ using the relation $\ket{\psi(t)}=\sum ^{N}_{m=1}
\psi_m(t)\ket{m}=e^{-iHt}\ket{\psi_{in}}$ where $m$ denotes the site
index that runs over all the $N$ sites i.e. $1,2 \ldots m, \ldots
N$. Also, selecting the `$u$' site yields similar results as the `$d$'
site case; thus, we do not show the results here.

\subsection{Zero disorder case}
We first study the system in the clean limit. We begin by
investigating the evolution of the particle density
where $p_m(t)=|\psi_m(t)|^2$ is the probability of site $m$ being
occupied at time $t$. When the initial state is chosen to be
$\ket{\psi_{in}}=\ket{c_k}$, the particle remains compactly localized
in two unit cells at all instances of time (see Fig.~\ref{fig2}(a)).
On the other hand, with the initial state taken to be
$\ket{\psi_{in}}=\ket{d_k}$, we observe that the particle becomes
compactly localized in three unit cells at all instances of time (see
Fig.~\ref{fig2}(c)). Also, the number of unit cells in which the
particle is compactly localized is robust with increasing system
size.

We next calculate the return probability, which is defined as:
\begin{equation}
R(t)=|\bra{\psi_{in}}\left.\psi(t)\right\rangle|^2.
\end{equation}
It is the probability of finding the particle in the initial state
after a time $t$. In the disorder-free limit, we have plotted the
return probability starting from both the initial states in the long
time limit $t=10^{9}$ in Fig.~\ref{fig2}(b) and Fig.~\ref{fig2}(d).
The spectrum is highly degenerate in the disorder-free limit, yielding
three energy levels, i.e. $E=\pm2, 0$. Since the return probability is
related to the level spacing of the energy levels, $R(t)$ shows
oscillatory behaviour~\cite{Kuno_2020, PhysRevA.100.043829}. We
conclude that the dynamics of the clean system is dependent on the
initial state.

\begin{figure*}
\centering
\stackunder{}{\includegraphics[width=0.75\textwidth]{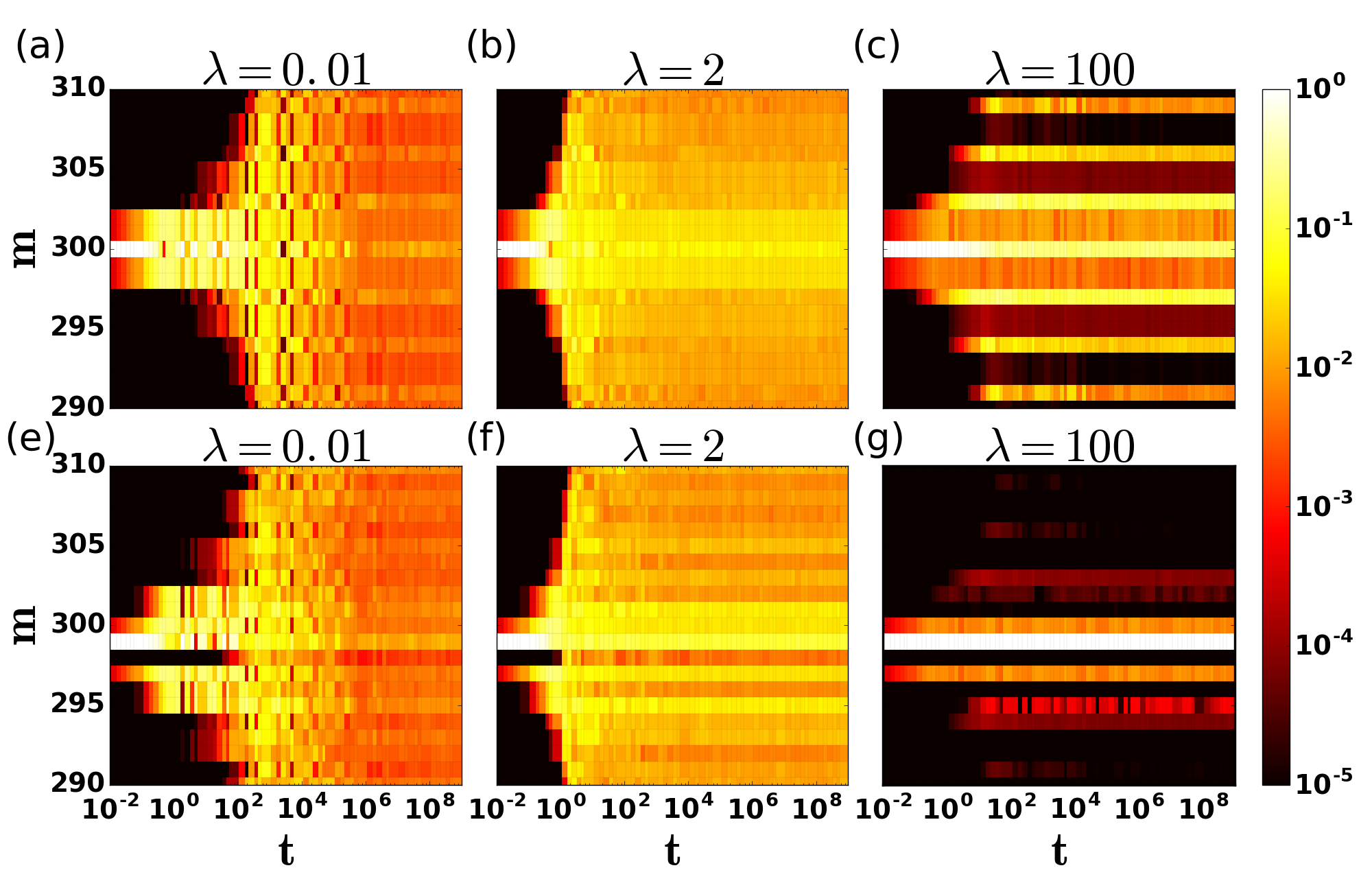}}\hspace{-3mm}
\stackunder{}{\includegraphics[width=0.25\textwidth]{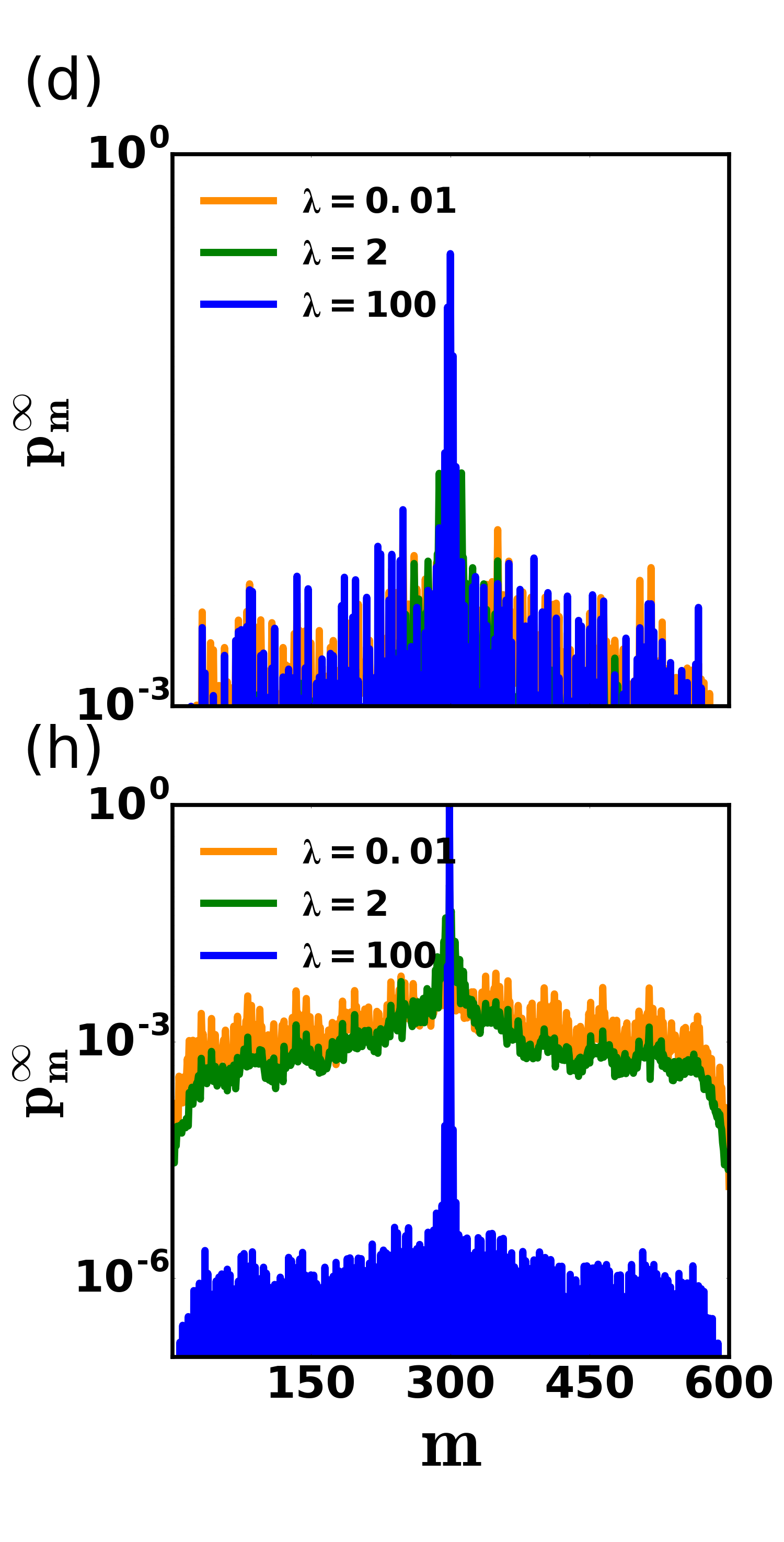}}
\caption{\label{fig4} In the antisymmetric case, the particle density (whose value is represented by a colour according to the code shown) as a function of time $t$, with $m$ denoting the site index, for a single particle initially at the $c-$site of the $100^{\text{th}}$ unit cell with increasing disorder strength (a)~$\lambda=0.01$, (b)~$\lambda=2$ and (c)~$\lambda=100$ and (d)~ particle density at $t=10^9$ for various strengths of disorder. For a single particle initially at the $d-$site of the $100^{\text{th}}$ unit cell, the particle density as a function of time $t$ for, (e)~$\lambda=0.01$, (f)~$\lambda=2$ and (g)~$\lambda=100$ and (h)~ particle density at $t=10^9$ for various strengths of disorder. Here system size is $N = 600$, and averaging over $100$ disorder realizations have been considered.}
\end{figure*}


\subsection{Symmetric disorder case}
We next consider the introduction of disorder in the symmetric configuration:
\begin{equation}
\zeta_{k}^{u}=\zeta_{k}^{d} \qquad \text{and} \qquad \zeta_{k}^{c}=0.
\label{eq3}
\end{equation}

We have previously observed~\cite{PhysRevB.106.205119} that 
in the single particle limit, although the degeneracy of all the flat bands is lifted,
the eigenstates are found to be compactly localized in two unit cells at all strengths of disorder.

For the initial state $\ket{\psi_{in}}=\ket{c_k}$, from the evolution
of the particle density, it can be observed that the
state is compactly localized over two unit cells in the low,
intermediate and high disorder regimes (see
Figs.~\ref{fig3}(a)--\ref{fig3}(c)). However, at higher disorder
$\lambda=100$, the site on which the particle is initially localized
shows a large occupation probability at all times, as indicated by the
central white patch in Fig.\ref{fig3}(c). Also, in the long time
limit, the return probability has finite magnitude $\approx 0.5$ for
$\lambda=0.01, 2$ and a magnitude close to unity for $\lambda=100$, as
shown in Fig.~\ref{fig3}(d). We then study the dynamics for the
initial state $\ket{\psi_{in}}=\ket{d_k}$. From the evolution of the
particle density, we observe that at all strengths of
disorder, the state is compactly localized over three unit cells (see
Figs.~\ref{fig3}(e)--\ref{fig3}(g)). The return probability in the
long time limit has a finite value at all disorder strengths, as shown
in Fig.~\ref{fig3}(h).
  
 On the introduction of a symmetric
disorder, the spectrum becomes dispersive, although the eigenstates
are compactly localized. We obtain non-degenerate energy levels, whose
magnitude depends on the disorder strength. As the return probability
involves the contribution of various energy levels through the time
evolution operator $U(t)=e^{-iHt}$, its periodicity is affected by the
various energy levels and the initial state.

There is a second way in which the symmetric disorder can be introduced wherein only the $c$
sites are perturbed:
\begin{equation}
\zeta_{k}^{u}=\zeta_{k}^{d}=0 \quad \text{and} \quad \zeta_{k}^{c}\neq 0.
\end{equation}
In this case, we know~\cite{PhysRevB.106.205119} that while the
degeneracy is broken for the upper and lower bands, the flat band at
$E=0$ remains robust even at higher disorder strengths. We have
checked that the single-particle dynamics within this scenario yields
qualitatively similar results as discussed above.

\subsection{Antisymmetric disorder case}
We next consider the application of the $AA$ potential in an antisymmetric manner, defined by
\begin{equation}
\zeta_{k}^{u}=-\zeta_{k}^{d} = \lambda\cos(2\pi k b+\theta_p) \quad \text{and} \quad \zeta_{k}^{c}=0.
\end{equation}

In the single-particle limit, we observed~\cite{PhysRevB.106.205119} that the tiniest of perturbation
lifted the degeneracy, and the eigenstates were no longer compactly
localized. Further, we also reported the existence of a central band with extended nonergodic
(multifractal) eigenstates separated from the Anderson localized states by a fractal mobility edge
$|E|<4/\lambda$~\cite{https://doi.org/10.48550/arxiv.2208.11930}.

From the evolution of the particle density, we observe
that for the initial state $\ket{\psi_{in}}=\ket{c_k}$ the
wavefunction spreads over the entire lattice with time $t$ at all
strengths of the disorder (see
Figs.~\ref{fig4}(a)--\ref{fig4}(c)). The same can be observed in the
long time limit $t=10^9$ (see Fig.~\ref{fig4}(d)), with occupation
probability $p_m^{\infty}$, spreading non-uniformly over the entire
space at all strengths of disorder, which is a signature of the
multifractal states, as observed in the phase diagram in the static
case~\cite{PhysRevB.106.205119}. The results are qualitatively the
same for the dynamics associated with the other initial state
$\ket{\psi_{in}}=\ket{d_k}$ as shown in
Figs.~\ref{fig4}(e)--\ref{fig4}(h). In both cases, in the higher
disorder regime $\lambda=100$, we observe that the site on which the
particle is initially localized shows a large occupation probability
in the long time limit (see Figs.~\ref{fig4}(d),(h)). For low and
intermediate disorder, we have also checked (results not shown here)
that the return probability $R(t)$ in the long time limit is of the
order of $O(10^{-2})$ due to the contribution of a large fraction of
multifractal eigenstates. At higher disorder strengths, it is $\approx
0.4$ for $\ket{\psi_{in}}=\ket{c_k}$ and $\approx 1$ for
$\ket{\psi_{in}}=\ket{d_k}$ owing to the presence of a large fraction
of localized eigenstates.


\begin{figure}[b]
\centering
\stackunder{}{\includegraphics[ width=7.5cm]{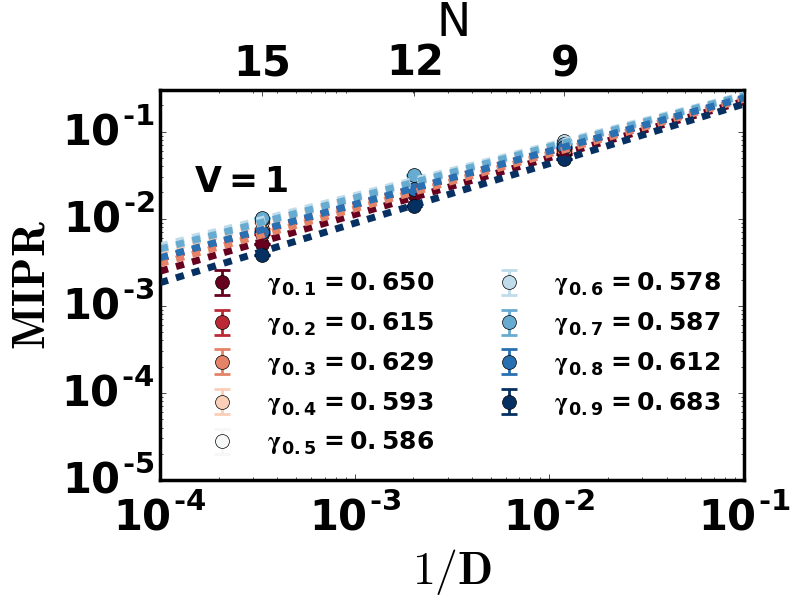}}
\caption{\label{fig5} MIPR averged over the eigenstates in the energy window $[\varepsilon-0.01,\varepsilon+0.01]$ with $1 / D$ where $\varepsilon=0.1,0.2,\ldots,0.9$, for a fixed interaction strength $V=1$. Here system sizes considered are $N=9, 12$ and $15$, and the filling fraction is fixed as $\nu=1/3$.}
\end{figure}

\section{INTERACTING DISORDER-FREE SYSTEM}\label{sec:level4}
In this section, we study the effects of the interaction $V$ on the
ABF diamond lattice in the zero disorder limit. We investigate the
properties of the eigenstates with the help of the many-particle
inverse participation ratio (MIPR) and the one-particle density matrix
(OPDM). We also explore the dynamics of the particle density,
entanglement entropy and return probability. For a system size $N$
with $N_p$ being the particle number, the dimension of the Hilbert
space is $D={N \choose N_p}$ and the filling fraction is represented
by $\nu=\frac{N_p}{N}$. Using exact diagonalization, we obtain the
many-body energy spectra ${E_i}$ and the normalized eigenstates
${\ket{\psi}_i}$, where $i=1,2,\ldots,D$.

Expanding a normalized eigenstate $|\Psi\rangle$ in the particle
number constrained space as $|\Psi\rangle=\sum_{i=1}^D C_i|i\rangle$,
we compute the many-particle inverse participation ratio (MIPR):
\begin{equation}
\mathrm{MIPR}=\sum_{i=1}^D\left|C_i\right|^4.
\end{equation}
For a perfect delocalized eigenstate MIPR $=O(1) / D$, while for an
extremely localized eigenstate, MIPR $=O(1)$. Here we study the
scaling of MIPR with $D$, using the relation MIPR $\propto
\frac{1}{D^{\gamma}}$. $\gamma$ is close to $0$ in the MBL phase,
while in a perfectly delocalized many-body phase $\gamma=1$ and in the
nonergodic many-body phase $0 < \gamma <
1$~\cite{PhysRevLett.123.180601}.


\begin{figure}
\centering
\stackunder{\hspace{-3.5cm}(a)}{\includegraphics[width=4.32cm]{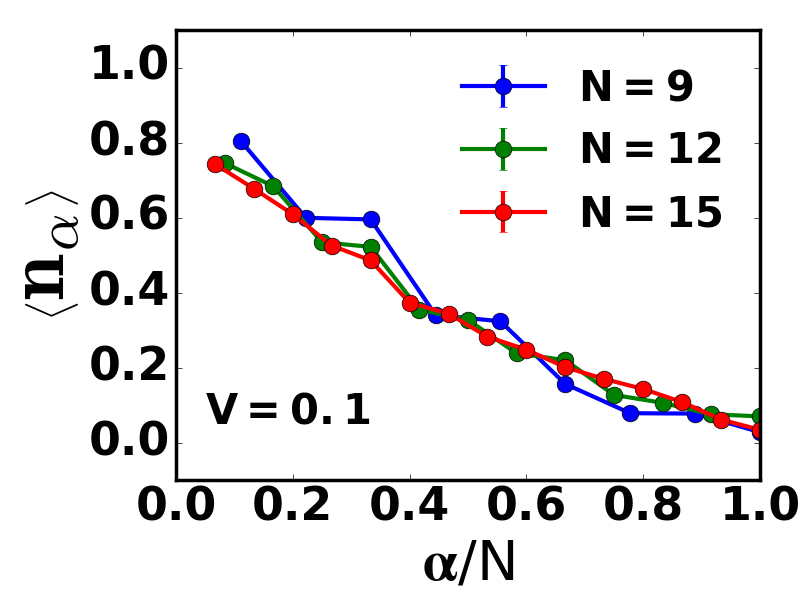}}\hspace{-1.2mm}
\stackunder{\hspace{-3.5cm}(b)}{\includegraphics[width=4.32cm]{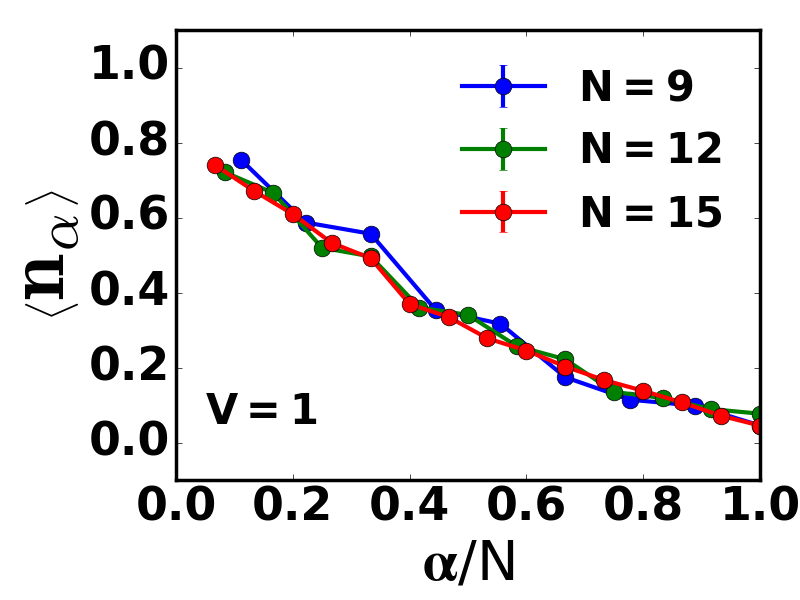}}
\vspace{-0.5cm}

\stackunder{\hspace{-3.5cm}(c)}{\includegraphics[width=4.32cm]{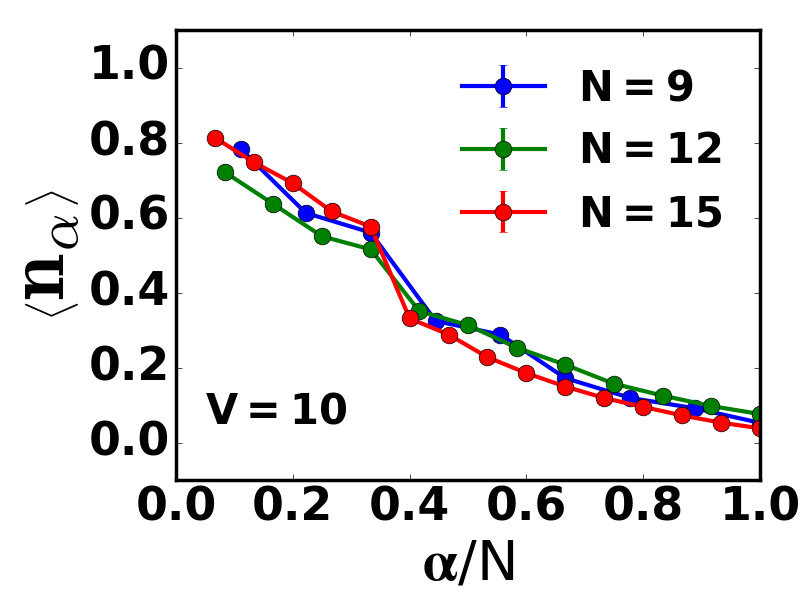}}\hspace{-1.2mm}
\stackunder{\hspace{-3.5cm}(d)}{\includegraphics[width=4.32cm]{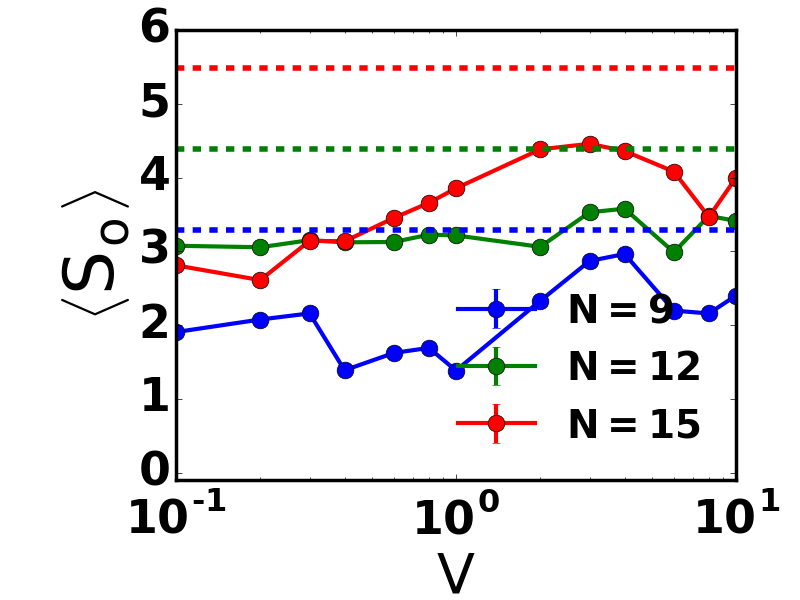}}
\caption{\label{fig6} Occupation spectrum $\left\langle n_\alpha \right\rangle$ with scaled index $\alpha/ N$ at fixed interaction strengths (a)~$V=0.1$, (b)~$V=1$, and (c)~$V=10$, for different system sizes $N=9, 12, 15$ and filling fraction $\nu=1/3$.  (d)~The average OPDM entropy $S_o$ with increasing interaction strength $V$. Dashed lines denote the maximal value of $S_o$. Averaging has been performed over the eigenstates in the energy window $\varepsilon=\left[0.54,0.57 \right]$.}
\end{figure}

In Fig.~\ref{fig5}, we fix the filling fraction $\nu=1/3$ and extract
$\gamma$ by increasing the system size $N$. Using the relation
$\varepsilon_i=\frac{E_i-E_1}{E_D-E_1}$, where $E_1$ and $E_D$ are the
ground state and maximum energy levels, respectively, the energy
levels are rescaled to lie within the range $0 \leq \varepsilon_i\leq
1$. We then study MIPR averaged over the states in the energy windows, which are specified as 
$[\varepsilon-0.01,\varepsilon+0.01]$, where $\varepsilon=0.1, 0.2,\ldots,0.9$ at $V=1$. From the
energy-resolved study, we observe that $0.57 \leqslant \gamma_\varepsilon
\leqslant 0.68$ over the entire energy spectrum, indicating the
existence of a nonergodic phase.

The localization characteristics of a many-body system can also be
explored with the help of the one-particle density matrix
(OPDM)~\cite{PhysRev.104.576,
  PhysRevB.96.060202,https://doi.org/10.1002/andp.201600356}. The OPDM
$\rho_o$ for any many-body eigenstate $\left|\Psi\right\rangle$ is
defined as:
\begin{equation}
{\left(\rho_o\right)}_{i j}=\left\langle\Psi\left|a_i^{\dagger} a_j\right| \Psi\right\rangle,
\label{eq9}
\end{equation}
where we have renamed the fermion operators at the various sites of
the different unit cells as $\left(u_1, d_1, c_1, u_2, d_2, c_2,
\ldots u_k, d_k, c_k \right)= \left(a_1, a_2, a_3 \ldots a_N \right)$
where $a_i^{\dagger}(a_i)$ creates(annihilates) a fermion on-site $i$
which runs from $i=1,2,\ldots,N$. A compact way to define these new operators is to simply write:
\begin{align}
  u_{k} &= a_{3(k-1)+1}\nonumber\\
  d_{k} &= a_{3(k-1)+2}\nonumber\\
  c_{k} &= a_{3(k-1)+3}
\end{align}
where $k = 1,2,\ldots,\frac{N}{3}$ runs over the unit cells. The
diagonalization of the OPDM results in a basis of single-particle
eigenstates called the natural orbitals
$\left|\phi_\alpha\right\rangle$, with $\alpha=1,2, \ldots, N$ and
their occupations (eigenvalues) denoted by $n_{\alpha}$:
\begin{equation}
\rho_o\left|\phi_\alpha\right\rangle=n_\alpha\left|\phi_\alpha\right\rangle.
\end{equation}
The trace of the OPDM is equal to the total number of particles in the
system $\text{tr}(\rho_o)=\sum_{\alpha=1}^N n_\alpha=N_p$, and the natural
orbitals are ordered with decreasing occupation: $n_1 \geq n_2 \geq
\ldots \geq n_N$.

These natural orbitals are localized in the MBL phase and delocalized
in the ergodic phase. This behaviour of the natural orbitals is a many
body effect since, without interactions, the natural orbitals of the
single-particle energy eigenstates are all localized. In a
non-interacting system, each many-body eigenstate
$\left|\Psi\right\rangle$ is a Slater determinant of $N_p$
single-particle states, with the occupation spectrum $n_\alpha=1$ for
all $\alpha \leq N_p$ and zero otherwise at any strength of
disorder. In the MBL phase, all the natural orbitals corresponding to
$\alpha \leq N_p$ remain almost fully occupied ($\left\langle
n_\alpha\right\rangle \approx 1$), while the others remain almost
unoccupied ($\left\langle n_\alpha\right\rangle \approx 0$), resulting
in a discontinuity $\delta n=n_{N_p+1}-n_{N_p}$ that is close to
unity. In the thermal phase, the occupations of all the orbitals
approach the mean filling fraction $\left\langle n_\alpha\right\rangle
\approx \nu$. In the ergodic phase, the occupation spectrum is
consistent with the eigenstate thermalization hypothesis, while in the
MBL phase, the occupations preserve a discontinuity at an emergent
Fermi edge.
\begin{figure}[b]
\centering
\stackunder{\hspace{-3.7cm}(a)}{\includegraphics[width=4.3cm]{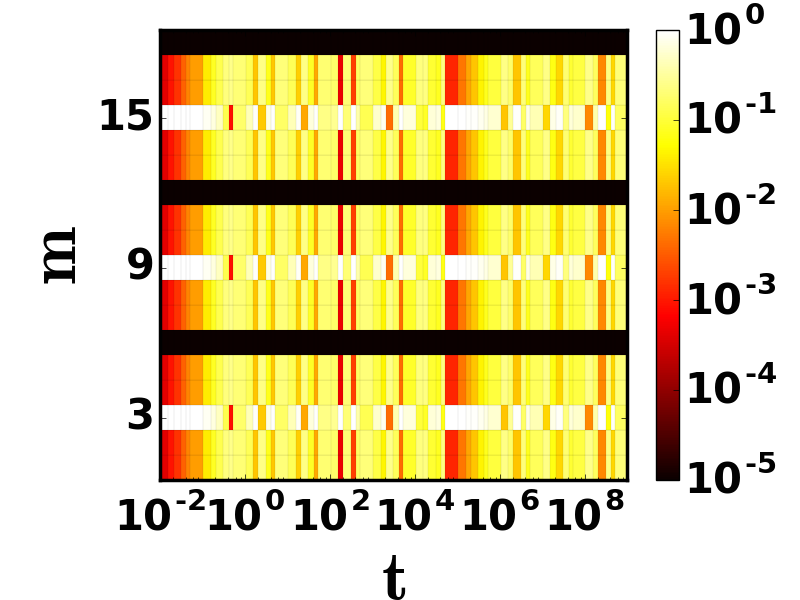}}\hspace{-1mm}
\stackunder{\hspace{-3.7cm}(d)}{\includegraphics[width=4.3cm]{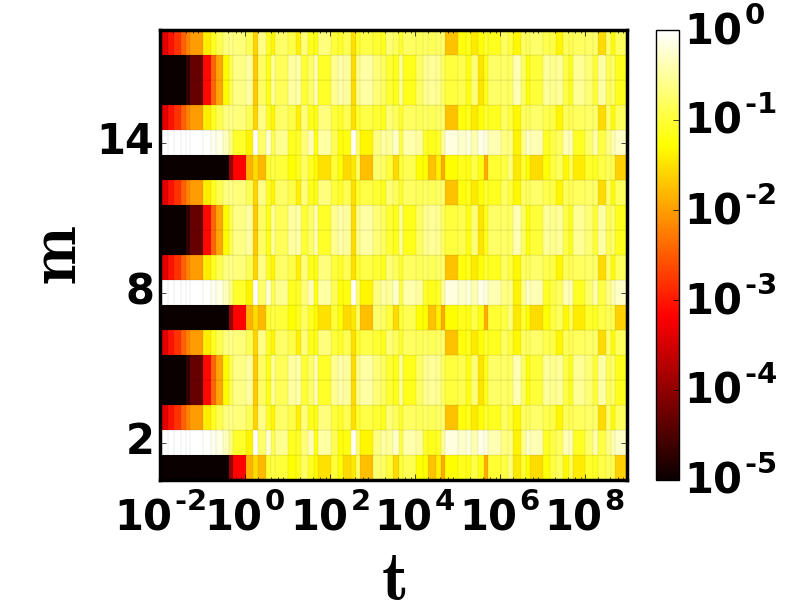}}
\vspace{-0.5cm}

\stackunder{\hspace{-3.7cm}(b)}{\includegraphics[width=4.3cm]{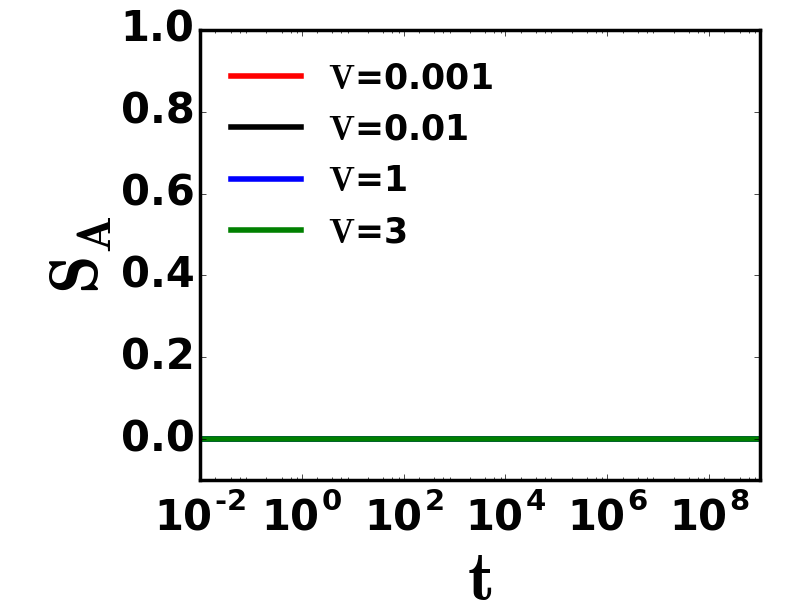}}\hspace{-1mm}
\stackunder{\hspace{-3.7cm}(e)}{\includegraphics[width=4.3cm]{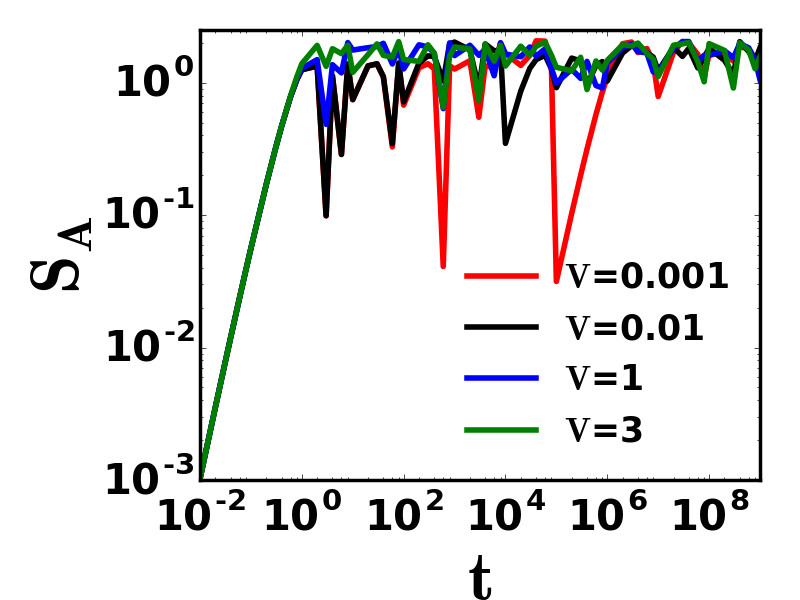}}
\vspace{-0.5cm}

\stackunder{\hspace{-3.7cm}(c)}{\includegraphics[width=4.3cm]{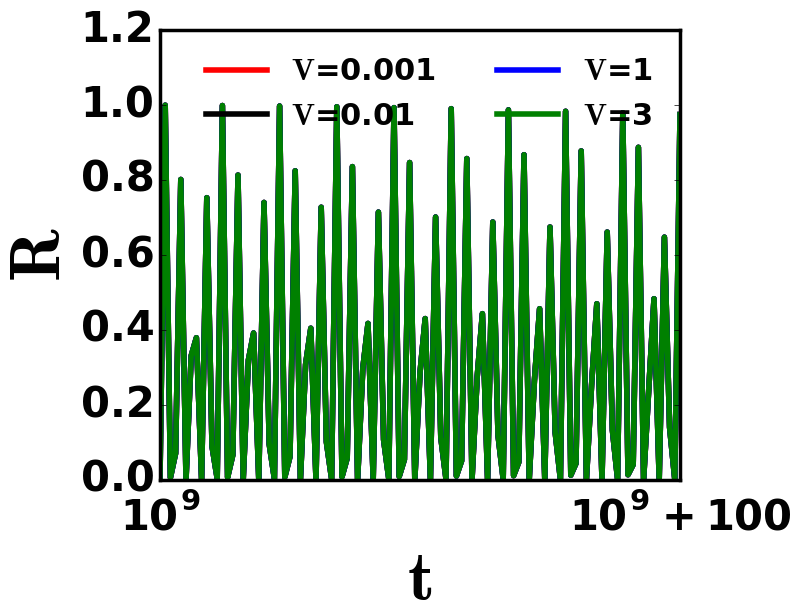}}\hspace{-1mm}
\stackunder{\hspace{-3.7cm}(f)}{\includegraphics[width=4.3cm]{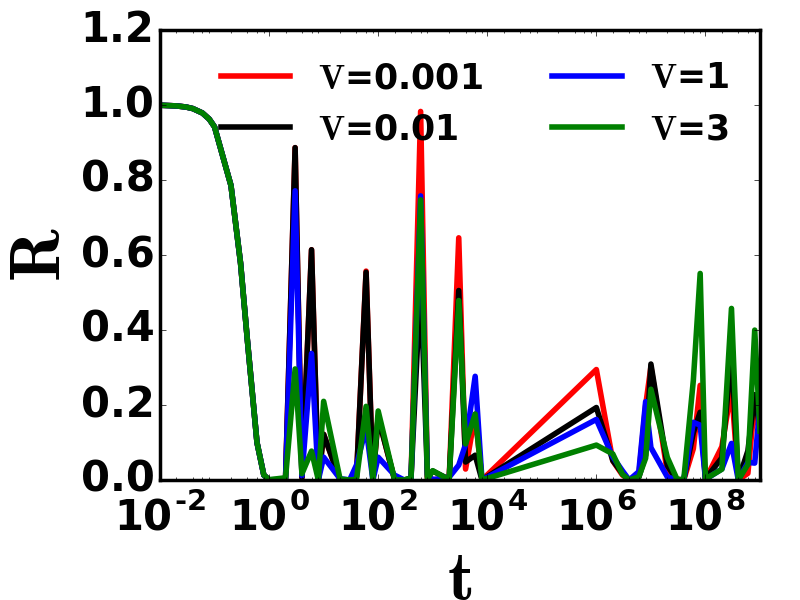}}
\caption{\label{fig7} For the initial state given by Eq.~\ref{i1},
  (a)~the particle density (whose value is represented by a colour
  according to the code shown) as a function of time $t$, where $m$ is
  the site index at interaction strength $V=1$, (b)~the entanglement
  entropy $S_A$ as a function of time $t$ for a subsystem of size
  $N_A=N/3$ and (c)~the return probability $R$ as a function of time
  $t$ for interaction strengths $V=0.001, 0.01, 1$ and $3$. (d-f)
  Corresponding plots for the initial state given by Eq.~\ref{i2}. For
  all cases, $N=18$ and $\nu=1/6$.}
\end{figure}

From the occupation spectrum, the one-particle occupation entropy can
be calculated as follows:
\begin{equation}
S_o=-\operatorname{tr} \rho_o \ln \rho_o=-\sum_\alpha n_\alpha \ln \left(n_\alpha\right).
\end{equation}
The one-particle occupation entropy is large and proportional to the
system size in the delocalized phase, corresponding to the volume law
of thermal states. In contrast, in the localized phase, it is close to
$0$. In the ergodic phase $\left\langle n_\alpha\right\rangle \approx
\nu$, hence for a filling fraction $\nu=1/3$ considered here, the
maximal value of $S_o$ will be $(N/3) \ln 3$. Thus in the thermal
phase, it corresponds to the volume law displayed by many-body
eigenstates, while it approaches $0$ in the localized phase.

In Figs.~\ref{fig6}(a)--\ref{fig6}(c), we have plotted the occupation
spectrum $\left\langle n_\alpha\right\rangle$ at different interaction
strengths $V=0.1, 1$ and $10$ and over a specific energy window
$\varepsilon=\left[0.54,0.57 \right]$. We observe that the occupations
are not close to $1$ or $0$, indicating the absence of MBL. Further
deep in the thermal phase, $\left\langle n_\alpha\right\rangle$ are
expected to become system size independent at the filling fraction
($\nu=1/3$ here), while splaying out on either side in a
characteristic system-size-dependent manner; here, we only see a
monotonic decrease with almost no system size dependence
throughout. We conclude that the presence of interaction in the ABF
diamond lattice results in a nonergodic phase - this is in agreement
with the results of MIPR. We also study the OPDM entropy $S_o$ (see
Fig.~\ref{fig6}(d)) for the states corresponding to the energies
$\varepsilon=\left[0.54,0.57 \right]$. We observe that $S_o$ does not
reach its maximal value (dashed lines in Fig.~\ref{fig6}(d)), nor does
it decrease to $0$ at any interaction strength $V$, and indicates
nonergodic behaviour.


In Section~\ref{sec:level3} from our discussion of the single-particle
dynamics, we have seen how the number of unit cells occupied by the
time-evolving state depends on the initial state. Here we study
many-body non-equilibrium dynamics with the help of particle
density, entanglement entropy and return probability. The study of
entanglement entropy~\cite{RevModPhys.82.277, LAFLORENCIE20161} serves
as a quantifier of localization in many-body systems. For the many-body state $\ket{\psi}$, one can calculate the density matrix $\rho=
\ket{\psi}\bra{\psi} $. The system is then divided into two parts, one
with $N_A$ number of sites and the other with $N_B=$ $N-N_A$
sites. The reduced density matrix (RDM) is calculated by tracing over
the subsystem $B$ as $\rho_A=\operatorname{Tr}_B(\rho)$, and the
entanglement entropy is given by $S_A=-\text{Tr}(\rho_A \ln \rho_A)$.

In order to understand the interplay of initial configuration and
interaction $V$, we consider two types of initial states for the
system size $N=18$ and a filling fraction $\nu=1/6$. In the first
case, we consider an initial state of the density wave type with particles
on $c-$sites of alternate unit cells~\cite{PhysRevResearch.2.043395}:
\begin{equation}
\ket{\psi_{in}^c}=\prod_{i=1}^{N/6} \hat{c}_{2i-1}^{\dagger}\ket{0}.
\label{i1}
\end{equation}
In the second type of initial state, the $d-$sites of alternate unit
cells are occupied:
\begin{equation}
\ket{\psi_{in}^d}=\prod_{i=1}^{N/6} \hat{d}_{2i-1}^{\dagger}\ket{0}.
\label{i2}
\end{equation}

Figs.~\ref{fig7}(a)--\ref{fig7}(c) show the dynamics starting from the
initial state given by Eq.~\ref{i1}. When the particles are arranged
in a manner such that $|h-l|\geq 2$ where $h, l$ are the unit cell
indices of any pair of particles, we observe distinct CLSs for each
particle as shown by the evolution of the particle density in
Fig.~\ref{fig7}(a). The particles show caging behaviour and remain
unaffected by the interactions here. The same can be observed from the
evolution of the entanglement entropy $S_A$, where $N_A=1/3$ in
Fig.~\ref{fig7}(b). We observe that at all interaction strengths,
there is zero entanglement between the two subsystems, indicating that
the compact localized states are unaffected by interaction strength
$V$. Also, we observe that the return probability shows perfect
oscillations (see Fig.~\ref{fig7}(c), where time axis is shown on a
linear scale to highlight the oscillations) in the long time limit
independent of the strength of interaction.

We then consider the initial configuration given by Eq.~\ref{i2},
corresponding to a $1/6$ filling fraction. However, the CLS
corresponding to the single particles spans over $3$ unit cells, as
shown in Fig.~\ref{fig2}(c). Consequently, from the evolution of the
particle density, we observe an overlap between the CLS belonging to
different particles in Fig.~\ref{fig7}(d). This suggests that 
interaction among the initially caged particles comes into play. The
same can also be observed from the evolution of $S_A$, where we plot
the entanglement by considering the subsystem size $N_A=1/3$. After an
initial transient till $t\approx1$, independent of the interaction
strength $V$, the entanglement saturates to a significant value
indicating a nonergodic phase. From Fig.~\ref{fig7}(f), we observe that
the return probability displays continual oscillations about a
non-zero mean value, although it does not reach $1$. A closer look at
this figure in a linear scale shows that the oscillations are
interaction dependent, mainly controlled by the interaction-dependent
gaps between the degenerate bands of the many-body energy
spectrum. These energy gaps are constant throughout the dynamics, and
hence, the associated terms present in the return probability do not
vanish due to phase randomization, giving rise to energy gap-dependent
fluctuations in return probability throughout the dynamics. This scenario is typical of a clean degenerate system
perturbed with many-body interactions~\cite{PhysRevResearch.2.043395}. This behaviour is also
consistent with the nonergodic phase argued from the previous
quantities.

\begin{figure}[b]
\centering
\stackunder{\hspace{-4cm}(a)}{\includegraphics[width=4.3cm]{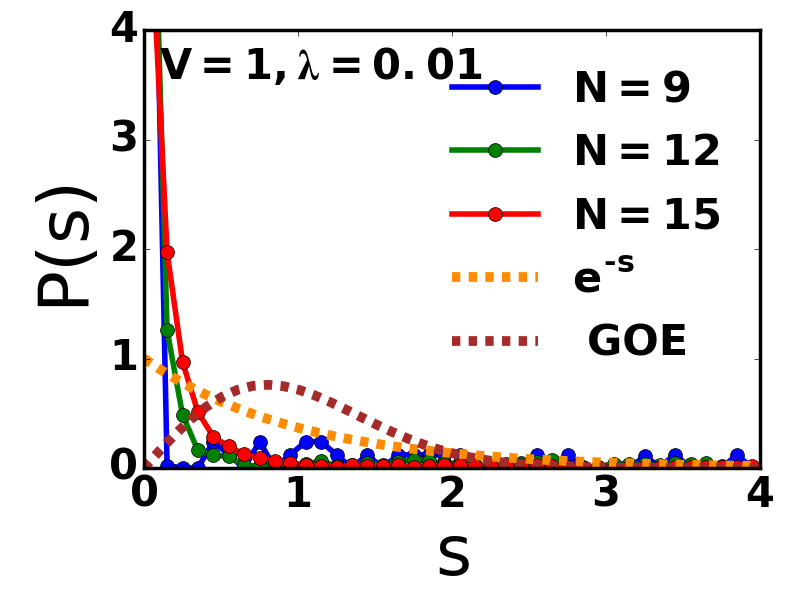}}\hspace{-2mm}
\stackunder{\hspace{-4cm}(b)}{\includegraphics[width=4.3cm]{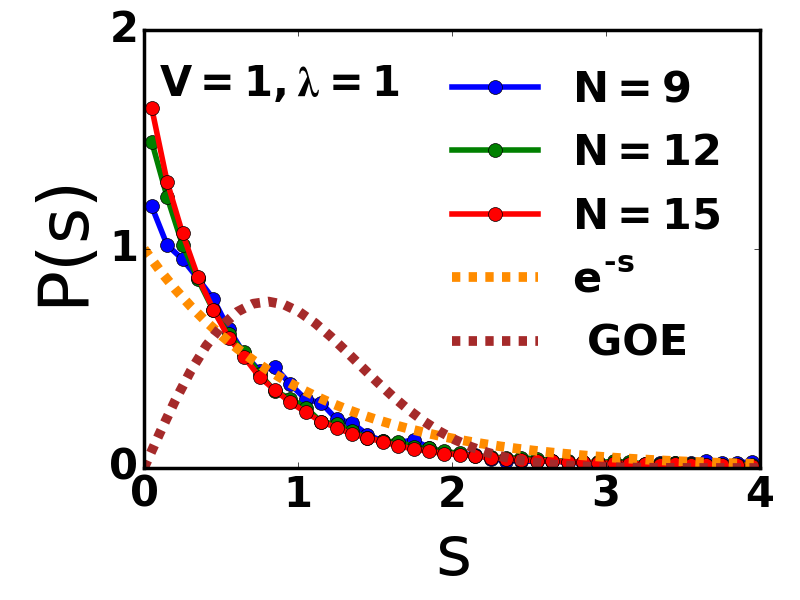}}
\caption{\label{fig8} In the symmetric case, level spacing distribution $P(s)$ with spacings $s$ at interaction strength $V=1$ for filling fraction $\nu=1/3$ at disorder strength (a)~$\lambda=0.01$ and (b)~$\lambda=1$.
The number of disorder realizations is $50$ for $N=15$ and $200$ for $N=9$ and $N=12$.}
\end{figure}
\section{INTERACTIONS AND SYMMETRIC  DISORDER}\label{sec:level5}
In this section, we study the interplay of symmetric disorder and
interactions. Specifically, we study the effects of the application of
$AA$ disorder in a symmetric manner where the disorder is introduced
on the $u$ and $d$ sites:
\begin{equation}
\zeta_{k}^{u}=\zeta_{k}^{d} \qquad \text{and} \qquad \zeta_{k}^{c}=0,
\end{equation}
in the presence of interactions. We first look at the eigenvalue and
eigenvector properties of this Hamiltonian, and then see how these
properties are reflected in a non-equilibrium dynamical setting.


\begin{figure*}
\centering
\stackunder{\hspace{-5.6cm}(a)}{\includegraphics[width=6cm]{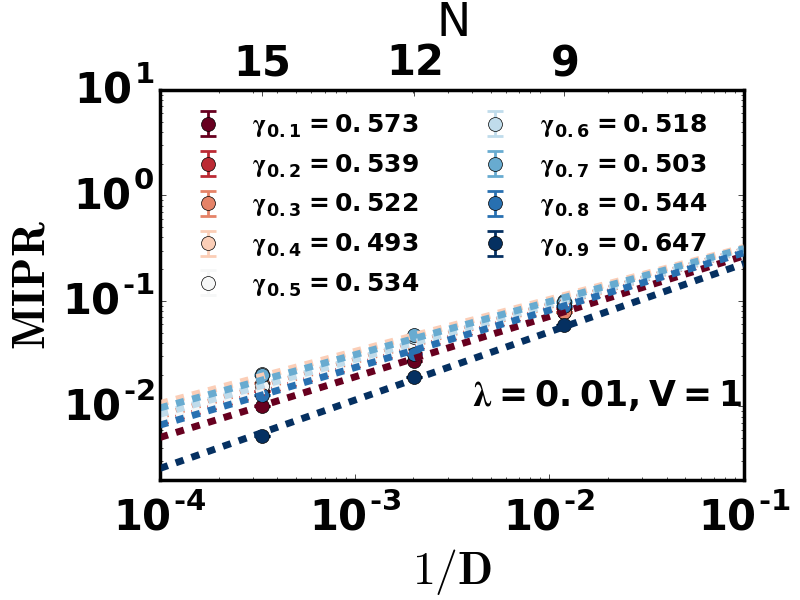}}\hspace{-2mm}
\stackunder{\hspace{-5.6cm}(b)}{\includegraphics[width=6cm]{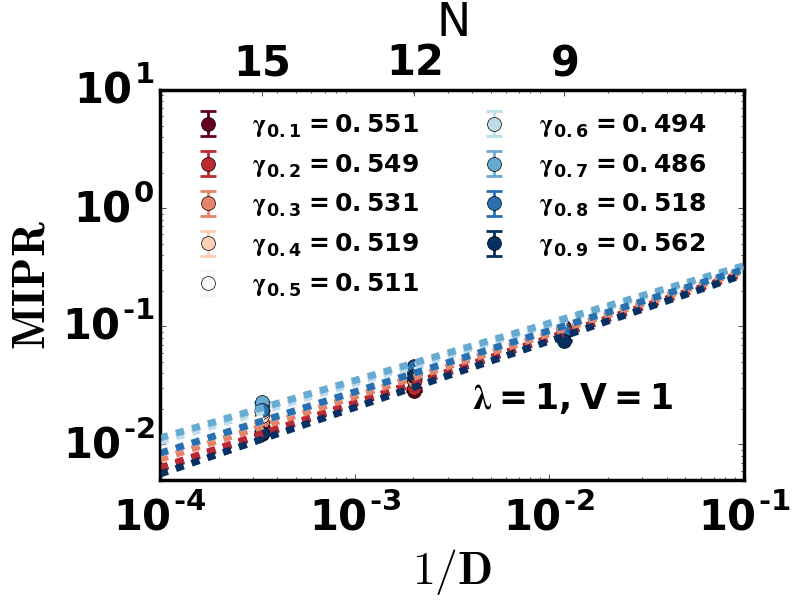}}\hspace{-2mm}
\stackunder{\hspace{-5.6cm}(c)}{\includegraphics[width=6cm]{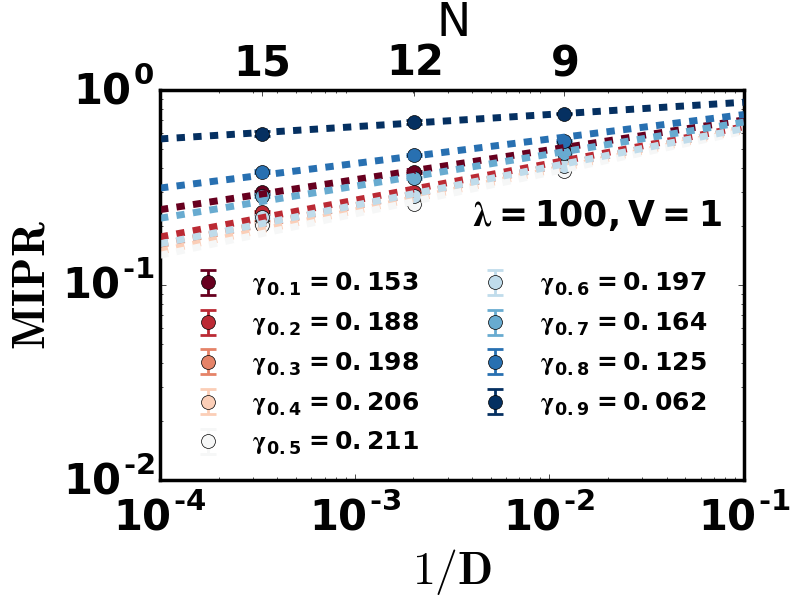}}\hspace{-2mm}
\caption{\label{fig9}In the symmetric case, MIPR averaged over states in the energy window $[\varepsilon-0.01,\varepsilon+0.01]$
with $1 / D$, where $\varepsilon=0.1,0.2,\ldots,0.9$ for a fixed interaction strength $V=1$ and disorder strength (a)~$\lambda=0.01$, (b)~$\lambda=1$ and (c)~$\lambda=100$. Number of disorder realizations are $400$, $200$, and $50$ for system sizes $N=9, 12$ and $15$, respectively and the filling fraction is $\nu=1/3$.}
\end{figure*}


\begin{figure}[b]
\centering
\stackunder{}{\includegraphics[ width=5.5cm]{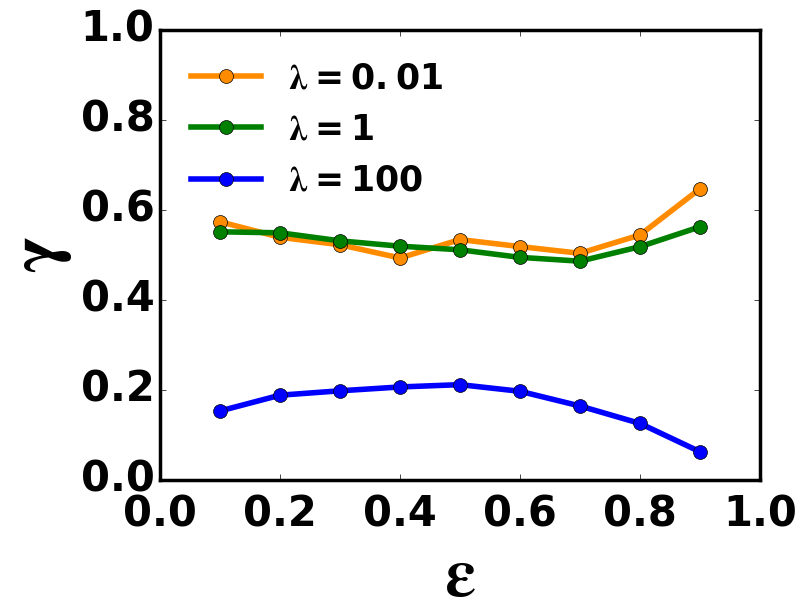}}
\caption{\label{fig10} The exponent $\gamma$ extracted from the energy resolved MIPR in Fig.~\ref{fig9} with rescaled energy $\varepsilon$ at interaction strength $V=1$ and disorder strength $\lambda=0.01, 1$ and $100$.}
\end{figure}

\subsection{Statics}\label{sec:level51}
We start by investigating the level spacing distributions $P(s)$. To
do this, the energy levels are arranged in ascending order, and the
consecutive spacings are obtained as $s_i=E_{i+1}-E_{i}$. A large
collection of such spacings is obtained with the aid of several
disorder realizations. Next, the $s_i$ are
unfolded~\cite{PhysRevE.81.036206} by dividing the original level
spacings by the mean level-spacing of the spectrum. We then study the
distribution of these scaled spacings. It is well known~\cite{PhysRevLett.110.084101} that when the
states involved are localized, the probability distribution of the
level spacings is Poissonian: $P(s)=e^{-s}$. On the other hand, for
delocalized states, the probability distribution of the level spacings
is Wigner-Dyson: $P(s)=\frac{\pi}{2}se^{-\frac{\pi}{4}s^2}$(GOE~\cite{PhysRevLett.110.084101}).

\begin{figure}[b]
\centering
\stackunder{\hspace{-3.5cm}(a)}{\includegraphics[width=4.32cm]{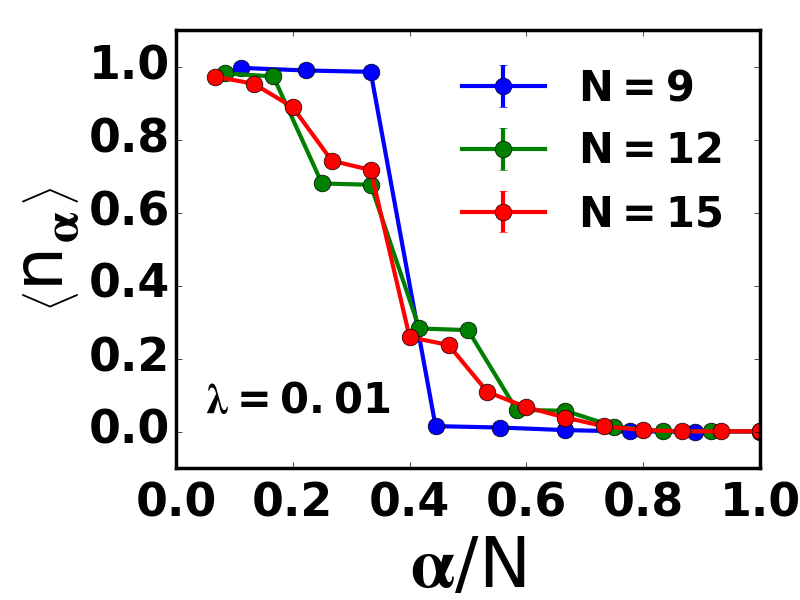}}\hspace{-1.2mm}
\stackunder{\hspace{-3.5cm}(b)}{\includegraphics[width=4.32cm]{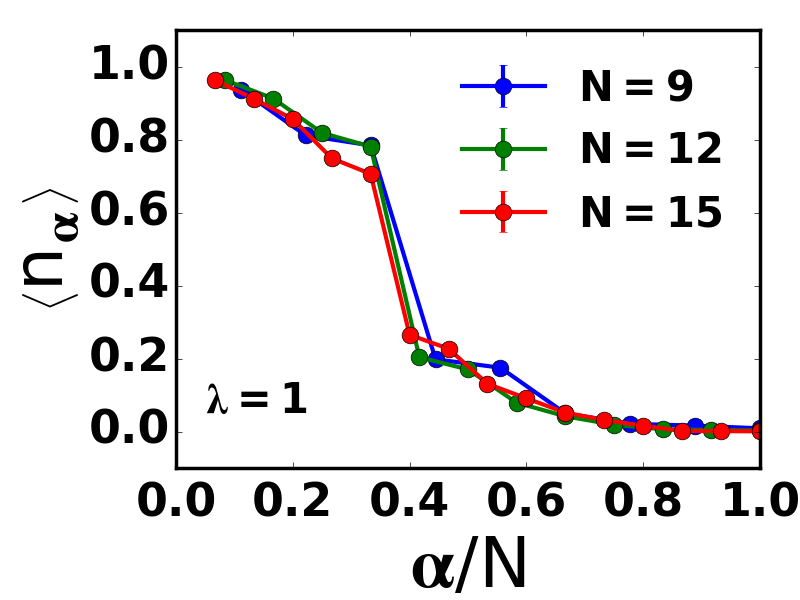}}
\vspace{-0.5cm}

\stackunder{\hspace{-3.5cm}(c)}{\includegraphics[width=4.32cm]{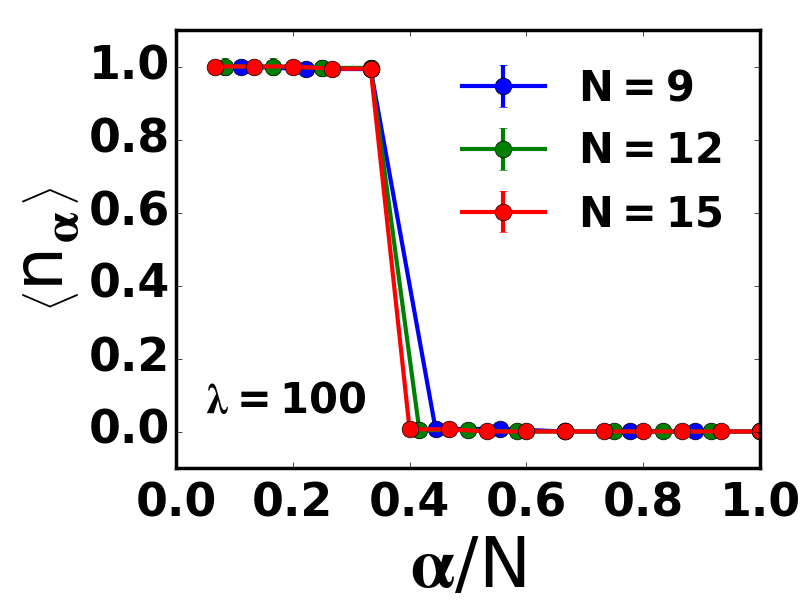}}\hspace{-1.2mm}
\stackunder{\hspace{-3.5cm}(d)}{\includegraphics[width=4.32cm]{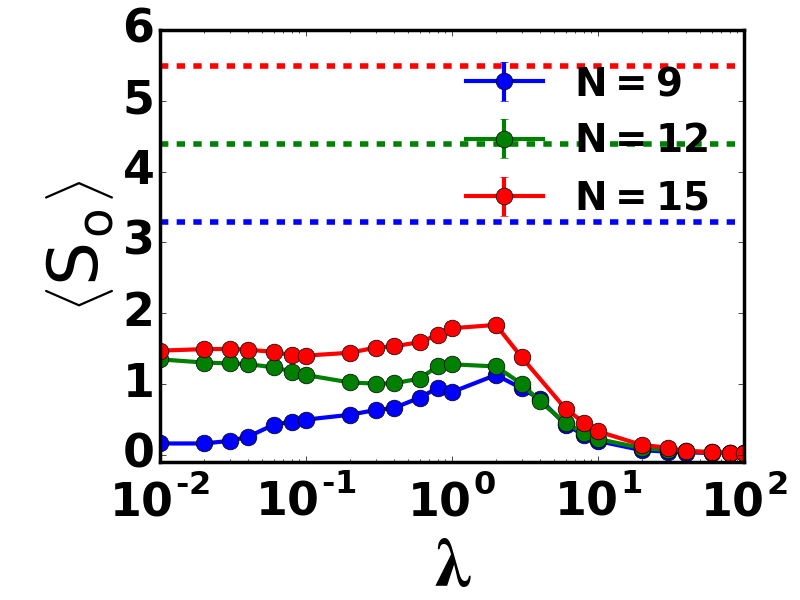}}
\caption{\label{fig11} Occupation spectrum $\left\langle n_\alpha \right\rangle$ with scaled index $\alpha/ N$ at fixed interaction strength $V=1$ and disorder strengths (a)~$\lambda=0.01$, (b)~$\lambda=1$, and (c)~$\lambda=100$, for different system sizes $N=9, 12, 15$ and fixed filling fraction $\nu=1/3$. (d)~The average OPDM entropy $S_o$ with increasing strength of disorder $\lambda$. Dashed lines denote the maximal value of $S_o$. Averaging has been performed over the eigenstates in the energy window $\varepsilon=\left[0.54,0.57 \right]$ and using $400,200$ and $50$ disorder realizations for system sizes $N=9, 12$ and $15$ respectively.}
\end{figure}

In the disorder-free case, the single-particle ABF diamond lattice
possesses massive degeneracy with only three energy levels. When
interactions are turned on for the disorder-free model, we observe
quasi-degeneracy as well as a large number of gaps in the spectrum.
On the application of disorder, while degenerate bands are observed in
the absence of interactions in the low disorder limit, many smaller
bands are observed when the interactions are also turned on. In the
case of high disorder, both in the presence and absence of
interactions, the spectrum displays quasi-degeneracy and many smaller
gaps, a behaviour typically observed in quasiperiodic
systems~\cite{PhysRevLett.123.025301}. This makes the level spacing
distribution not a reliable tool for studying localization
characteristics~\cite{PhysRevLett.122.040606} in the low and high
disorder regime. In the intermediate disorder regime, these effects
are minimized due to the interplay of flat bands and disorder.
Fig.~\ref{fig8} shows the probability distribution of the level
spacing at interaction strength $V=1$ for a fixed filling fraction
$\nu=1/3$ and different disorder strengths $\lambda$. We observe that
the spacing distribution is neither GOE nor does it show a perfect fit
to the Poisson distribution both at $\lambda=0.01$ (see
Fig.~\ref{fig8}(a)) and at $\lambda=1$ (see Fig.~\ref{fig8}(b)). The
states are neither ergodic nor localized in the low and intermediate
disorder regimes.

\begin{figure}
\centering
\stackunder{}{\includegraphics[ width=0.5\textwidth]{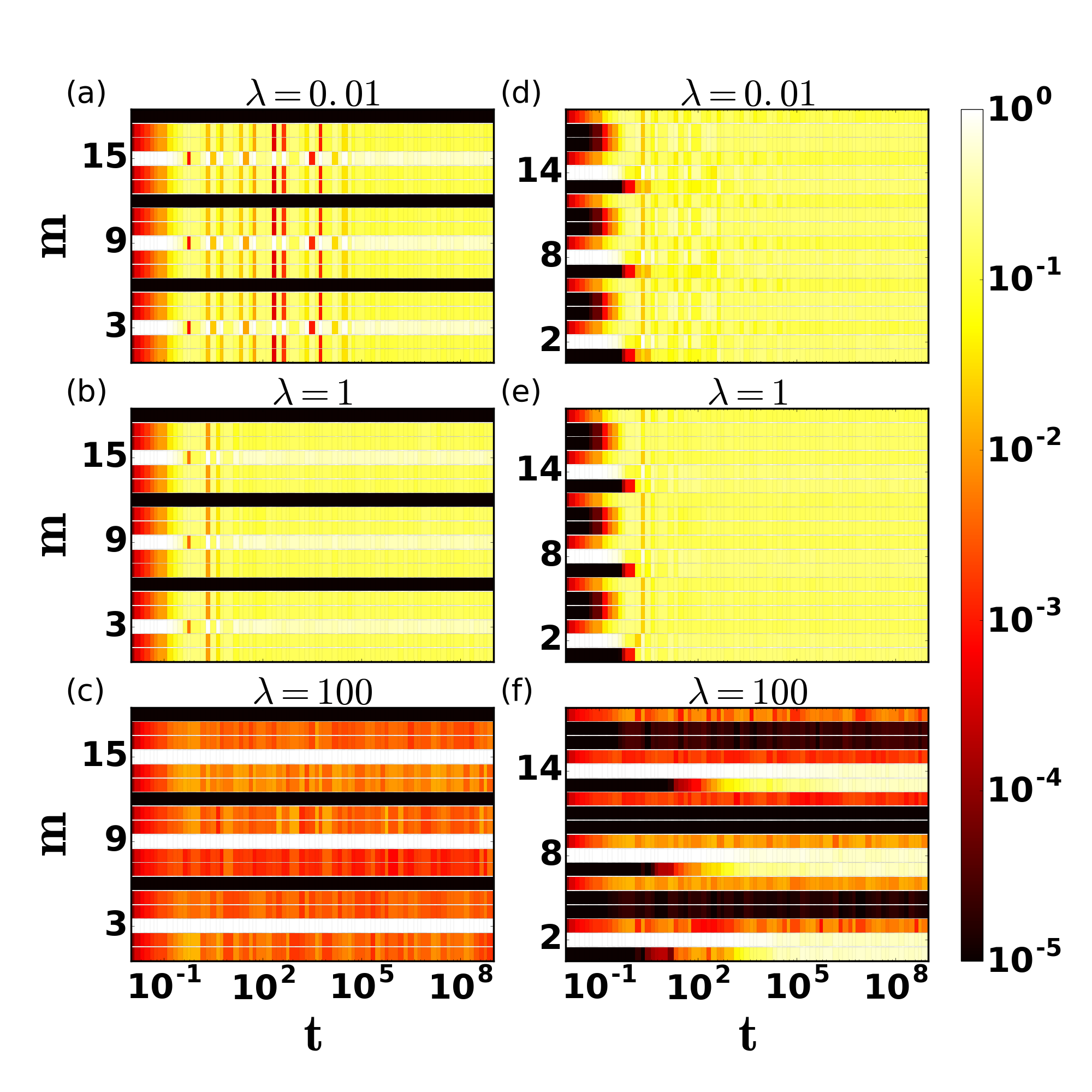}}
\caption{\label{fig12} In the symmetric case, particle density (whose value is represented by a colour according to the code shown) as a function of time $t$, where $m$ is the site index, for the initial state given by Eq.~\ref{i1} for interaction strength $V=1$ and increasing disorder strengths (a)~$\lambda=0.01$, (b)~$\lambda=1$, (c)~$\lambda=100$. Corresponding plots show the evolution of the particle density (d)--(f)~for the initial state given by Eq.~\ref{i2}. $N=18$, $\nu=1/6$ and $100$ disorder realizations have been considered for all cases.}
\end{figure}

\begin{figure}[b]
\centering
\stackunder{\hspace{-4.0cm}(a)}{\includegraphics[width=4.2cm]{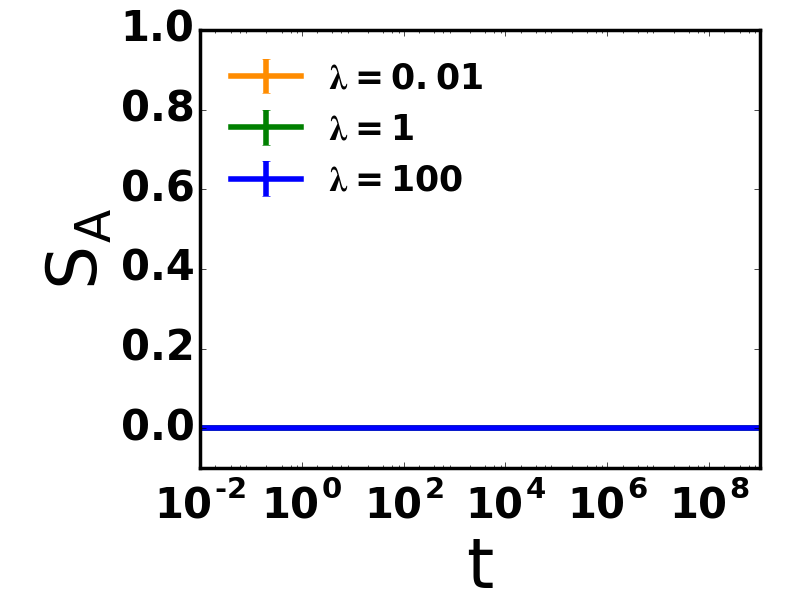}}
\stackunder{\hspace{-4.0cm}(b)}{\includegraphics[width=4.2cm]{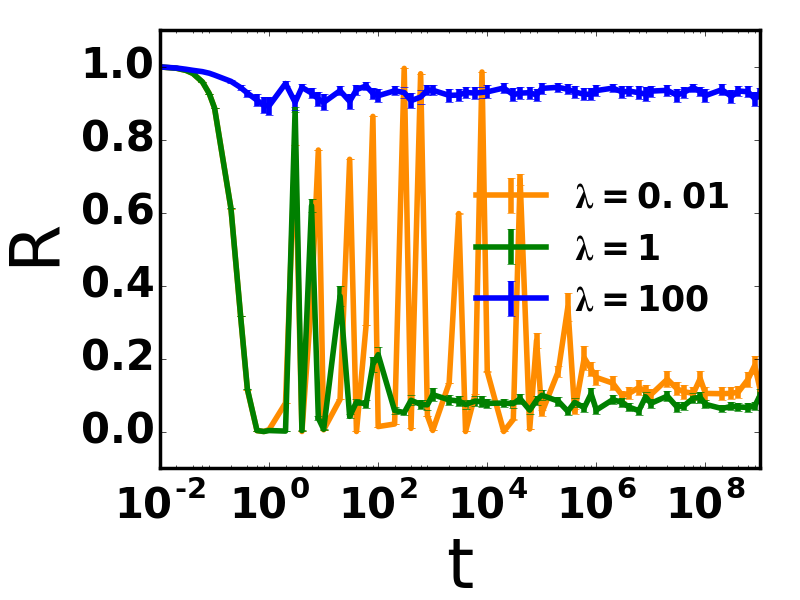}}
\vspace{-0.4cm}

\stackunder{\hspace{-4.0cm}(c)}{\includegraphics[width=4.2cm]{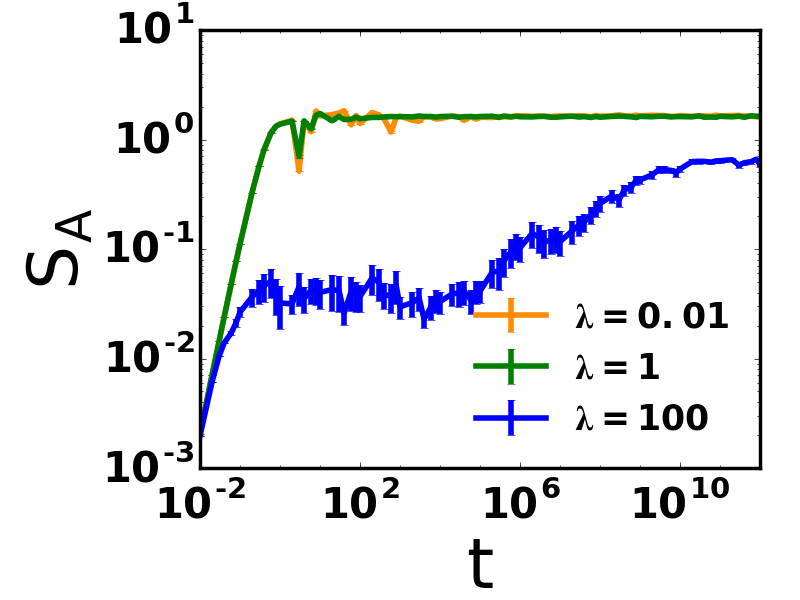}}
\stackunder{\hspace{-4.0cm}(d)}{\includegraphics[width=4.2cm]{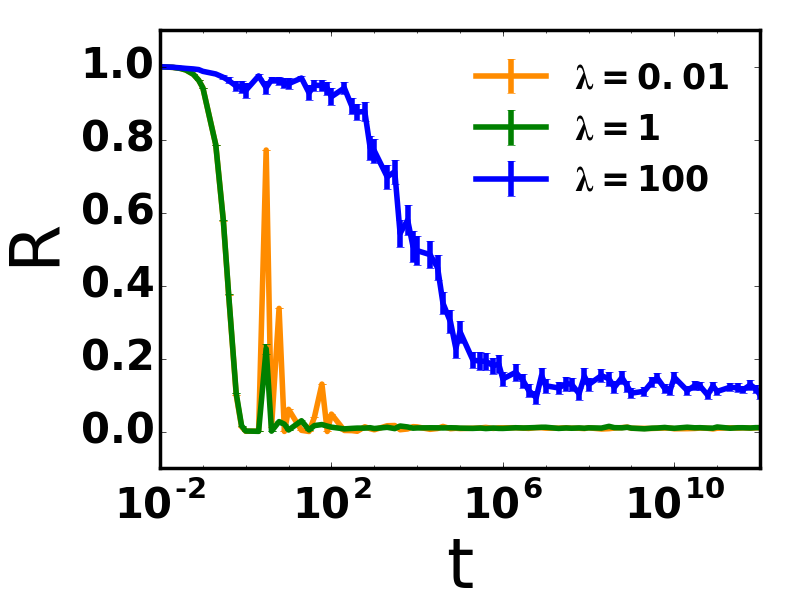}}
\caption{\label{fig13}In the symmetric case, (a) entanglement entropy
  $S_A$ for a subsystem of size $N_A=N / 3$ and (b) return probability
  $R$ as a function of time $t$ for the initial state given by
  Eq.~\ref{i1}. (c) Entanglement entropy $S_A$ for a subsystem of size
  $N_A=N / 3$ and (d) return probability $R$ as a function of time
  $t$ for the initial state given by Eq.~\ref{i2}. The interaction
  strength is fixed as $V=1$ with increasing disorder strengths
  $\lambda$. For all plots, $N=18, \nu=1 / 6$ and number of disorder
  realizations is $100$.}
\end{figure}

\begin{figure}
\centering
\stackunder{\hspace{-3.8cm}(a)}{\includegraphics[ width=4.2cm]{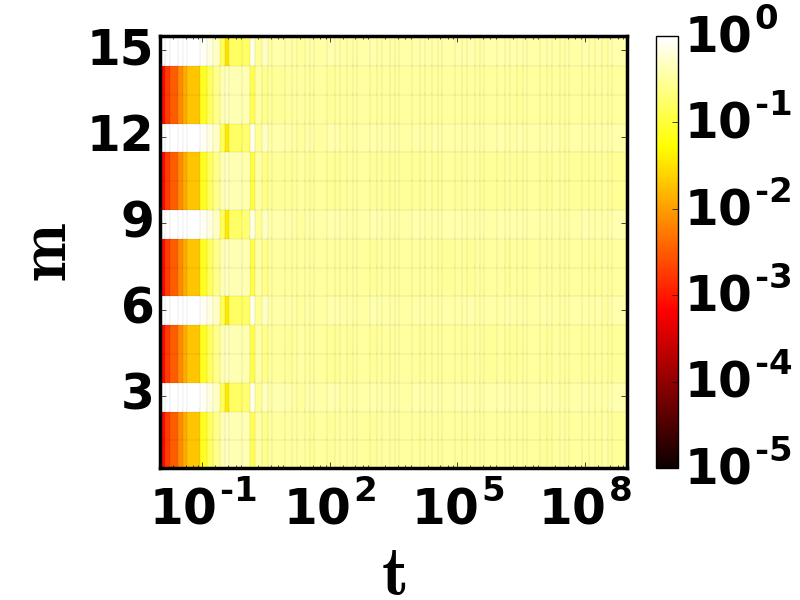}}
\stackunder{\hspace{-3.8cm}(b)}{\includegraphics[width=4.2cm]{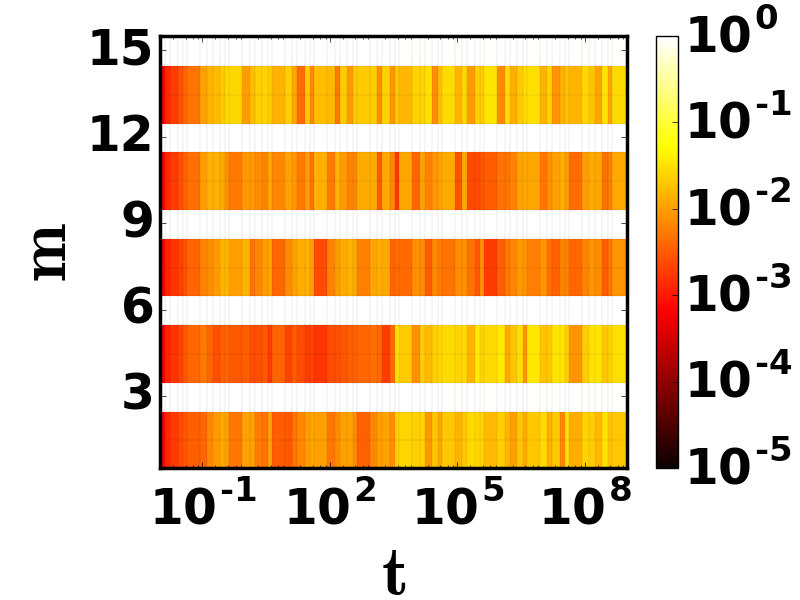}}
\vspace{-0.3cm}

\stackunder{\hspace{-3.8cm}(c)}{\includegraphics[width=4.2cm]{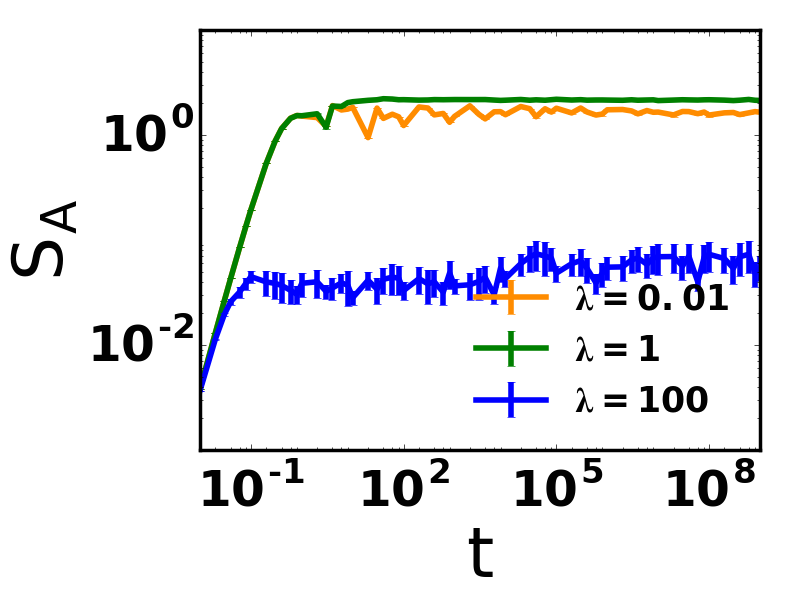}}
\stackunder{\hspace{-3.8cm}(d)}{\includegraphics[width=4.2cm]{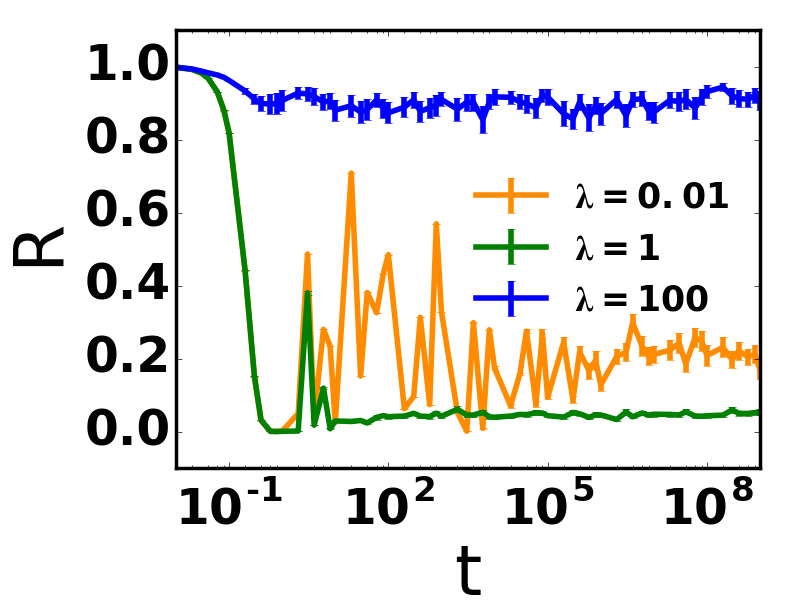}}

\vspace{-0.3cm}
\stackunder{\hspace{-3.8cm}(e)}{\includegraphics[width=4.2cm]{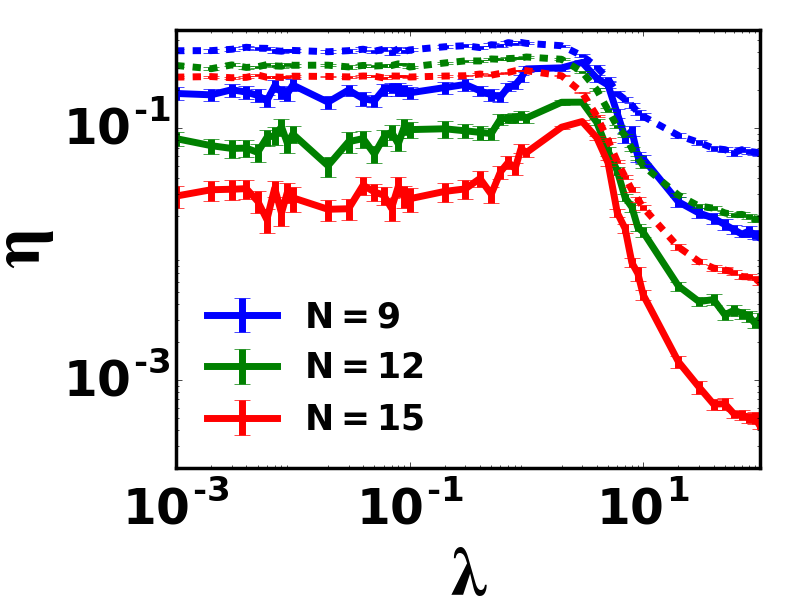}}
\stackunder{\hspace{-3.8cm}(f)}{\includegraphics[width=4.2cm]{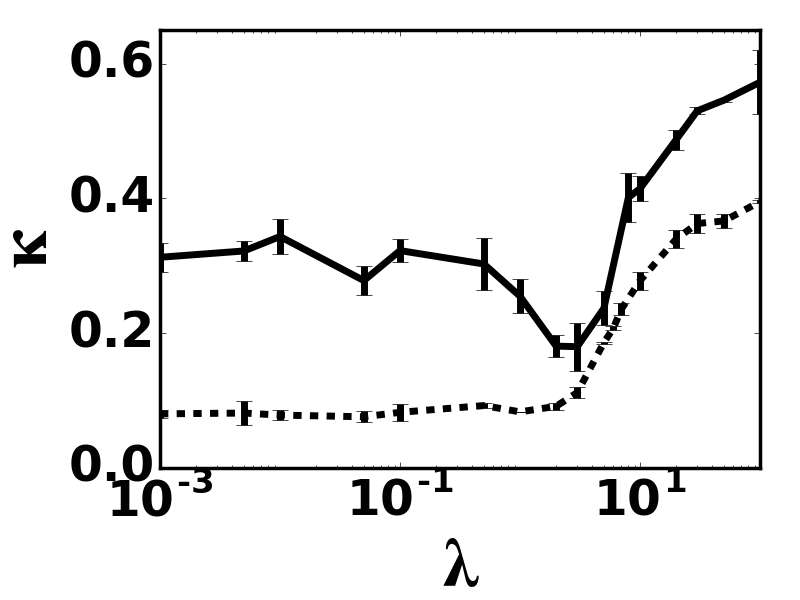}}
\caption{\label{fig14}The particle density (whose value is represented
  by a colour according to the code shown) as a function of time $t$,
  where $m$ is the site index for the initial state given by
  Eq.~\ref{i3} and for disorder strengths (a) $\lambda=1$ and (b)
  $\lambda= 100$. (c) Entanglement entropy $S_A$ for a subsystem of
  size $N_A=N / 3$ and (d) return probability $R$ as a function of
  time $t$. Here the interaction strength is fixed to $V=1$, and the
  system size is $N=15$, with filling fraction $\nu=1 / 3$ and
  averaging has been done over $50$ disorder realizations. (e) In the
  long time limit $t=10^9$, the NPR $\eta$ as a function of disorder
  strength $\lambda$ for various system sizes $N$, solid lines
  correspond to the initial state given by Eq.~\ref{i3} while dashed
  lines correspond to the initial state given by Eq.~\ref{i4}. (f) The
  scaling exponent $\kappa$ as a function of disorder strength
  $\lambda$. Here $V=1$ and $\nu=1 / 3$, and the number of disorder
  realizations are at least $50$ for all the system sizes.}
\end{figure}

As discussed above, quasi-degeneracy and gaps in the low and high
disorder limits yield inconclusive results when we study eigenvalue
properties. We will now discuss MIPR at a fixed interaction strength
$V=1$ and various disorder strengths $\lambda$ as shown in
Fig.~\ref{fig9}. For a fixed filling fraction $\nu=1/3$, we extract
the exponent $\gamma$ by averaging the MIPR over the states belonging
to the energy window $[\varepsilon-0.01,\varepsilon+0.01]$, where
$\varepsilon=0.1, 0.2...0.9$. In the low disorder regime
$\lambda=0.01$ (see Fig.~\ref{fig9}(a)), $\gamma \approx 0.6$ over the
entire energy spectrum indicating a nonergodic phase.  We observe
similar nonergodic behaviour in the intermediate regime $\lambda=1$
as shown in Fig.~\ref{fig9}(b). However, in the high disorder regime
with $\lambda=100$ (see Fig.~\ref{fig9}(c)), the exponent has a
significantly lower value $\gamma\approx0.1$ which is a signature of
MBL-like behaviour. In Fig.~\ref{fig10}, we have plotted the exponent
$\gamma$ at different disorder strengths and observe that it shows 
consistent behaviour over the entire spectrum.

We also study the OPDM here, with the help of the occupation spectrum
$\left\langle n_\alpha\right\rangle$ at different disorder strengths
$\lambda=0.01, 1$ and $100$ (see Figs.~\ref{fig11}(a)--\ref{fig11}(c))
and over a specific energy window $\varepsilon=\left[0.54,0.57
  \right]$. At low and intermediate disorder strengths, the occupation
spectrum falls monotonically with practically no dependence on system
size and with no signature of the thermal value $\left\langle
n_\alpha\right\rangle = \nu =1/3$; it also does not quite reach close
to $0$ or $1$ either indicating nonergodic behaviour. However, in the
high disorder regime, $\lambda=100$, it reaches close to $0$ and $1$,
indicating localized behaviour of the single-particle states and hence
an MBL-like phase. The OPDM entropy $S_o$ is also consistent with the
above inferences. While it is quite far from its thermal value
(represented by dashed lines in Fig.~\ref{fig11}(d)) in the low and
intermediate disorder regimes, it shows system size independence and
goes close to $0$ in the MBL-like phase.

\subsection{Nonequilibrium dynamics}
Next, we study the many-body non-equilibrium dynamics with the help of particle density, entanglement entropy and return probability. For the initial state given by
Eq.~\ref{i1}, we observe the evolution of the particle density
that the CLSs corresponding to distinct
particles remain isolated and unaffected by the interaction strength
$V$ as shown in Figs.~\ref{fig12}(a)--\ref{fig12}(c) for disorder
strengths $\lambda=0.01, 1$ and $100$ which results in caging. For
the initial state given by Eq.~\ref{i2}, the amplitude
corresponding to different CLSs overlap, and the interaction comes
into play. From the time evolution of the particle density (see
Figs.~\ref{fig12}(d)--~\ref{fig12}(f)), we observe that the compact
localized nature is no longer sustained. While nonergodic behaviour
is observed at low and intermediate disorder, at higher disorder
strength ($\lambda=100$), it is comparatively less nonergodic in the
long time limit.

We also study the entanglement entropy and return probability dynamics
for both the initial configurations as shown in Fig.~\ref{fig13}. In
the case of the initial state corresponding to Eq.~\ref{i1}, for a
subsystem of size $N_A=1/3$, we observe that at all disorder
strengths, $S_A\approx0$ (see Fig.~\ref{fig13}(a)) which supports the
observation of caging from the particle density. From the return
probability dynamics (see Fig.~\ref{fig13}(b)), with increasing
disorder strengths, and for $V=1$, we observe that while in the low and
intermediate disorder regimes, $R(t)$ has a finite value, in the
higher disorder regime, it approaches unity. For the second initial
state (Eq.~\ref{i2}), in the low and intermediate disorder regimes
after the transient, we observe a sub-diffusive regime, followed by
$S_A$ reaching saturation at a value of the order of the thermal value
as shown in Fig.~\ref{fig13}(c). In the high disorder regime, we
observe a logarithmically slow growth which eventually saturates to a
sub-thermal value. The same can be observed from the evolution of
return probability (see Fig.\ref{fig13}(d)). The dynamics of return
probability supports the results observed from the particle density
(see Figs.~\ref{fig12}(d)--~\ref{fig12}(f)) and entanglement entropy
with $R(t)$ close to $0$ in the low and intermediate disorder regimes
and with a finite magnitude in the high disorder regime.

In Fig.~\ref{fig14}, we study the dynamics when the initial state is
of the density wave type with particles on $c-$sites of every unit
cell, i.e. with filling fraction $\nu=1/3$:
\begin{equation}
\ket{\psi_{in}}=\prod_{i=1}^{N/3} \hat{c}_{i}^{\dagger}\ket{0}.
\label{i3}
\end{equation}

\begin{figure}[b]
\centering
\stackunder{\hspace{-4cm}(a)}{\includegraphics[width=4.1cm]{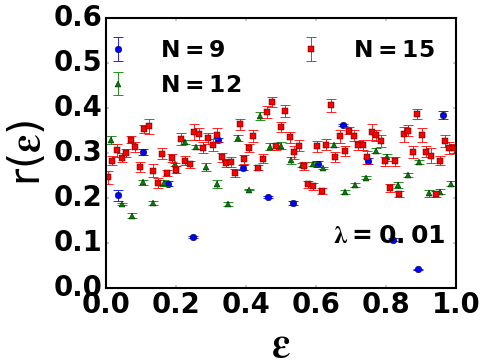}}\hspace{-1mm}
\stackunder{\hspace{-4cm}(b)}{\includegraphics[width=4.1cm]{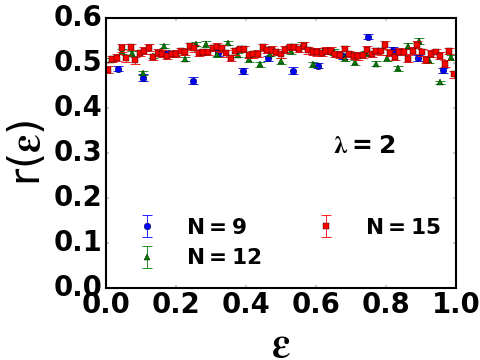}}
\caption{\label{fig15} In the antisymmetric case, energy-resolved gap-ratio as a function of the fractional eigenstate index $\epsilon$ for disorder strength (a)~$\lambda=0.01$ and (a)~$\lambda=2$. The number of disorder realizations is $500, 400$, and $50$ for system sizes $N=9, 12$ and $15$, respectively.}
\end{figure}

\begin{figure*}
\centering
\stackunder{\hspace{-5.6cm}(a)}{\includegraphics[width=6cm]{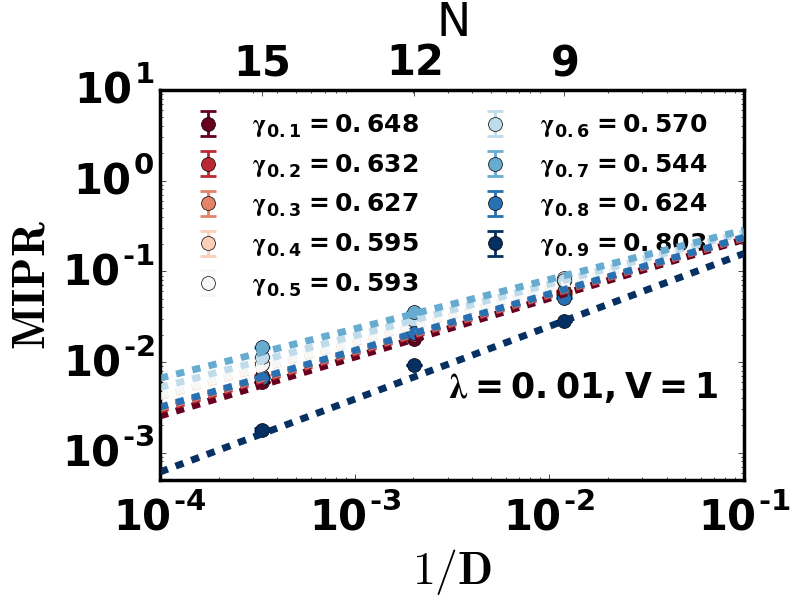}}\hspace{-2mm}
\stackunder{\hspace{-5.6cm}(b)}{\includegraphics[width=6cm]{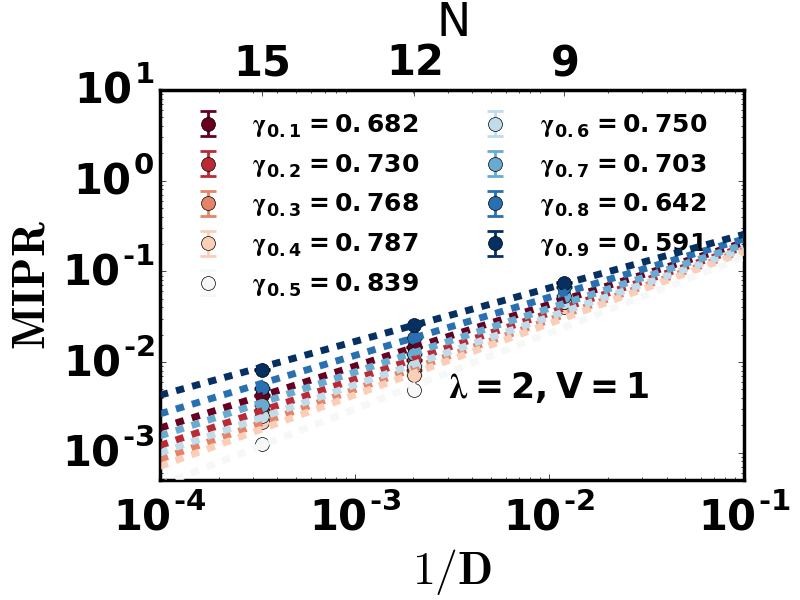}}\hspace{-2mm}
\stackunder{\hspace{-5.6cm}(c)}{\includegraphics[width=6cm]{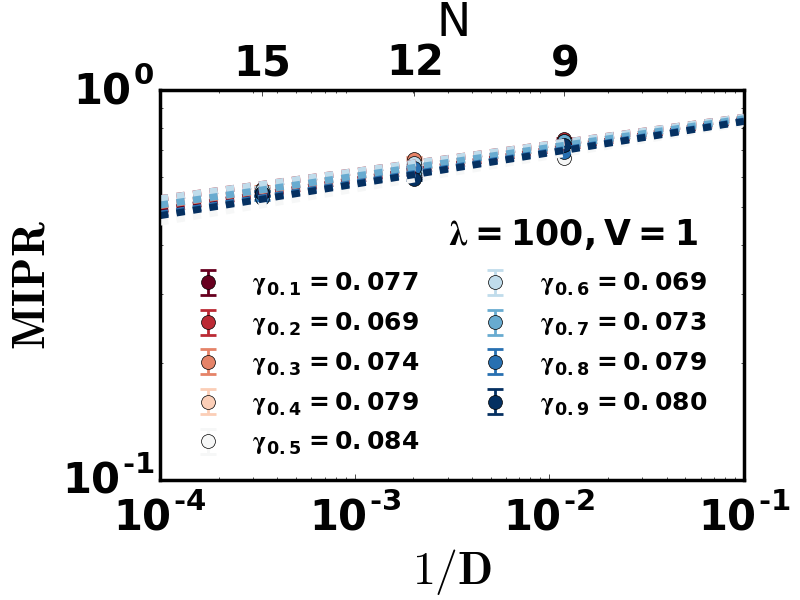}}\hspace{-2mm}
\caption{\label{fig16}~In the antisymmetric case, MIPR averaged over states in the energy window $[\varepsilon-0.01,\varepsilon+0.01]$ with $1 / D$, where $\varepsilon=0.1,0.2,\ldots,0.9$ for a fixed interaction strength $V=1$ and disorder strength (a)~$\lambda=0.01$, (b)~$\lambda=2$ and (c)~$\lambda=100$. Number of disorder realizations are $400$, $200$, and $50$ for system sizes $N=9, 12$ and $15$, respectively and the filling fraction is $\nu=1/3$.}
\end{figure*}

The evolution of the particle density for the interaction strength
$V=1$ and disorder strengths $\lambda=1$ and $100$, is shown in
Fig.~\ref{fig14}(a)--(b). While at low (not shown here) and
intermediate disorder, we observe that the particle density spreads
uniformly over all the sites, at high disorder strength $\lambda=100$,
the particle density is significantly localized over the initially
occupied sites, indicating MBL-like behaviour. We also study the
dynamics of entanglement entropy $S_A$ as shown in Fig.~\ref{fig14}(c)
at $V=1$. After the initial transient, $S_A$ shows a subdiffusive
growth in the low and intermediate disorder regimes and saturates to a
large value indicating delocalization in the many-body system. The
high disorder regime shows a logarithmic growth with time $t$,
saturating to a much lower magnitude compared to the thermal value
indicating MBL-like behaviour. The return probability dynamics is
shown in Fig.~\ref{fig14}(d). For low disorder, it saturates to a
finite value, indicating nonergodic behaviour, while in the
intermediate disorder regime, the magnitude is much smaller (but $\neq
0$), indicating that the phase has a higher nonergodic tendency. At
higher disorder strengths, it is close to unity, which is a signature
of MBL-like behaviour.

We further study the normalized participation ratio
(NPR)~\cite{PhysRevB.87.134202}, in the long time limit ($t=10^9$) to understand
the many-body phases. We consider two types of initial states with filling fraction
$\nu=1/3$, one given by Eq.~\eqref{i3} and another density wave type state with
particles on $d-$sites of every unit cell:
\begin{equation}
\ket{\psi_{in}}=\prod_{i=1}^{N/3} \hat{d}_{i}^{\dagger}\ket{0}.
\label{i4}
\end{equation} 

For any time evolved many-body state,
$|\Psi(t)\rangle=\sum_{i=1}^D C_i(t)|i\rangle$, the NPR is given as:
$$
\eta=\frac{1}{D \sum_i\left|C_i\right|^4}.
$$ In the long time limit $(t\rightarrow \infty)$, $\eta$ is
independent of system size $N$ in the ergodic phase; in contrast, it
decays exponentially with the system size in the localized
phase~\cite{PhysRevB.87.134202}. In Fig.~\ref{fig14}(e), we study the
dependence of NPR $\eta$ on the disorder strength $\lambda$, fixed
interaction strength $V=1$ and increasing system sizes. Here the solid
lines correspond to the initial state given by Eq.~\eqref{i3} while
the dashed lines correspond to the initial state given by
Eq.~\eqref{i4}. We observe that in both cases, $\eta$ is system size
dependent at all strengths of disorder $\lambda$, indicating the
absence of the thermal phase. Further, the exponent $\kappa$ can be
extracted at various disorder strengths using the relation $\eta
\propto e^{-\kappa N}$. In Fig.~\ref{fig14}(f), we plot $\kappa$ with
increasing disorder strength and observe that for both initial
states, in the low and intermediate disorder regime $0<\kappa<0.5$,
indicating nonergodic behaviour. However, at higher disorder
strength, it reaches near $0.5$, which is a sign of many-body
localizatione~\cite{PhysRevB.87.134202}.
 
For the second type of symmetric configuration, when the disorder is only considered on the $c-$site:
\begin{equation}
\zeta_{k}^{u}=\zeta_{k}^{d}=0 \quad \text{and} \quad \zeta_{k}^{c}\neq 0
\end{equation}
we observe that the results are qualitatively similar to the one discussed above.

\section{INTERACTIONS AND ANTISYMMETRIC  DISORDER}\label{sec:level6}
In this section, we study the interplay of antisymmetric disorder and
interactions. Specifically, we consider the antisymmetric application
of the $AA$ disorder on the $u$ and $d$ sites:
\begin{equation}
\zeta_{k}^{u}=-\zeta_{k}^{d} \qquad \text{and} \qquad \zeta_{k}^{c}=0,
\end{equation}
in the presence of interactions. We first study the eigenvalue and
eigenvector properties, and then investigate them in a non-equilibrium
dynamical setting as well.
\begin{figure}[b]
\centering
\stackunder{}{\includegraphics[ width=5.5cm]{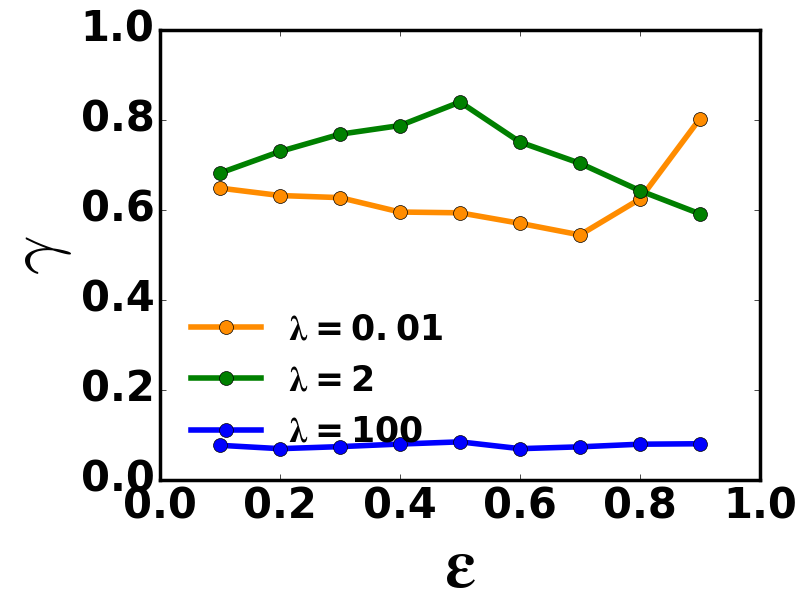}}
\caption{\label{fig17}~The exponent $\gamma$ extracted from the energy resolved MIPR in Fig.~\ref{fig16} with rescaled energy $\varepsilon$ at interaction strength $V=1$ and disorder strength $\lambda=0.01, 2$ and $100$. }
\end{figure}

\subsection{Statics}
We begin by analyzing the eigenvalue properties with the aid of the
level-spacing ratio $r_{av}$~\cite{PhysRevLett.110.084101}, defined
as:
\begin{eqnarray}
r_{av}=\bigg\langle\frac{1}{N-2}\sum\limits_{i=1}^{N-2}\frac{\min[s_{i},s_{i+1}]}{\max[s_{i},s_{i+1}]}\bigg\rangle.
\label{eq_level}
\end{eqnarray}
Here the energies $E_i$'s are first organized in ascending order,
which are used to obtain the energy level-spacings
$s_i=E_{i+1}-E_i$. The braces in Eq.~\ref{eq_level} represent the
average over disorder realizations. In the delocalized and localized
phases, $r_{av}$ is expected to be approximately $0.528$ and $0.386$,
respectively~\cite{PhysRevLett.110.084101}. In Fig.~\ref{fig15}, we
study the energy-resolved level spacing ratio by dividing the
many-body energy spectrum into several equal segments and calculating
the local average of the level-spacing ratio for each segment of the
spectrum. While in the low disorder regime (see Fig.~\ref{fig15}(a))
$r_{av}\approx 0.3$ indicating a mixed non-ergodic phase,
$r_{av}\approx0.52$ for $\lambda=2$ (see
Fig.~\ref{fig15}(b)) suggests a thermal-like phase. As discussed in
Section~\ref{sec:level51}, in the high disorder regime, the spectrum
displays quasi-degeneracy and many gaps; we do not show the results
here as they are inconclusive.
\begin{figure}[b]
\centering
\stackunder{\hspace{-3.5cm}(a)}{\includegraphics[width=4.3cm]{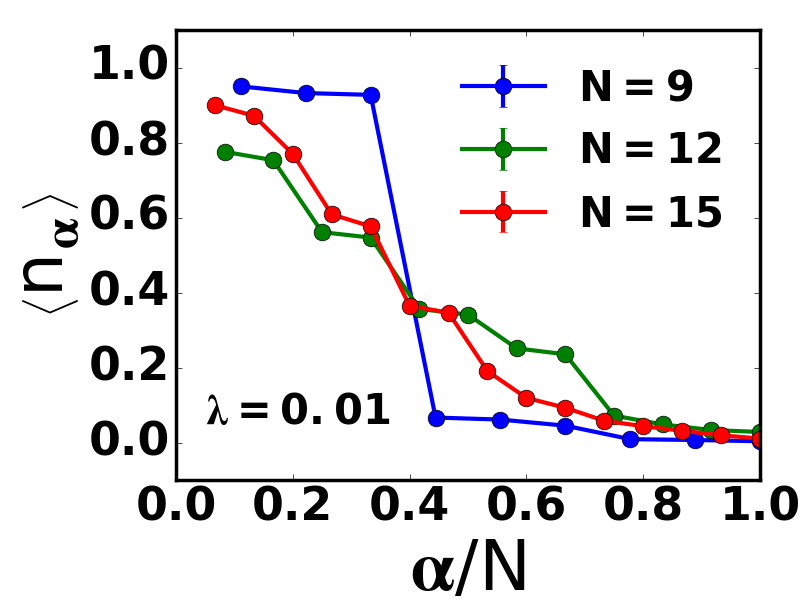}}\hspace{-1mm}
\stackunder{\hspace{-3.5cm}(b)}{\includegraphics[width=4.3cm]{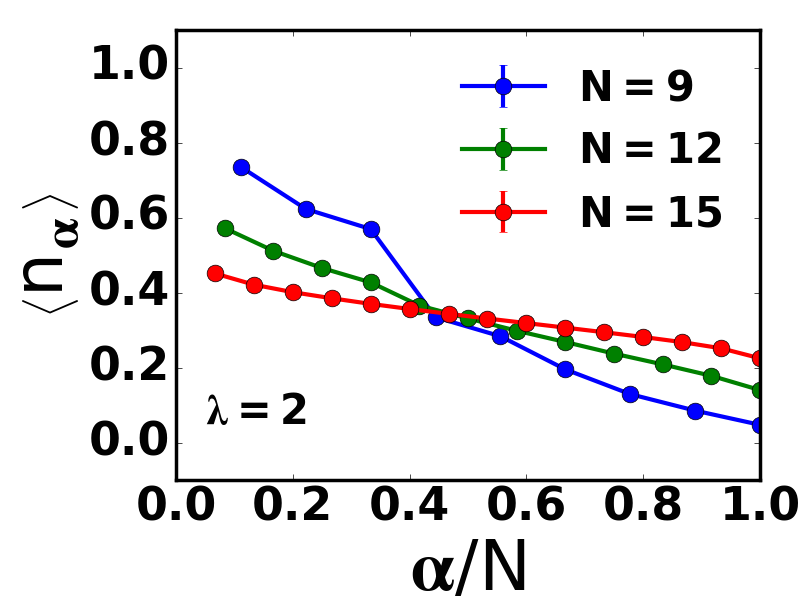}}
\vspace{-0.5cm}

\stackunder{\hspace{-3.5cm}(c)}{\includegraphics[width=4.3cm]{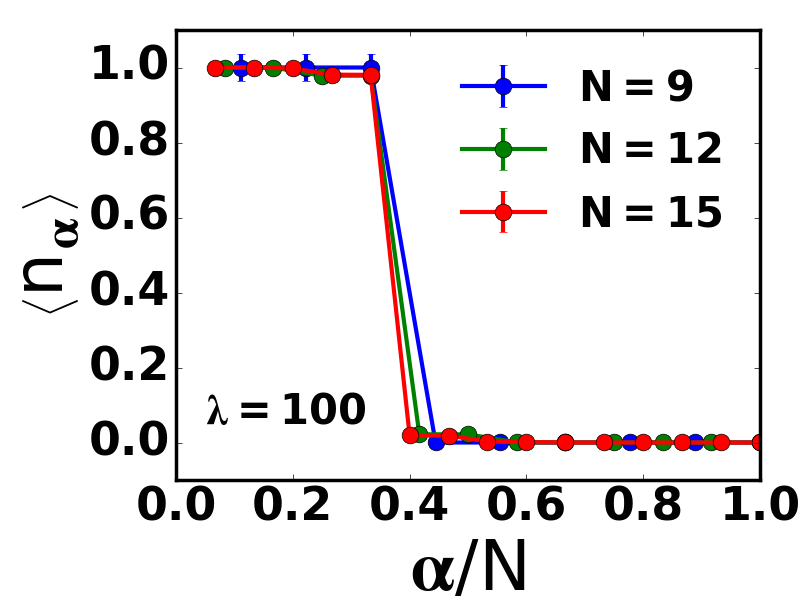}}\hspace{-1mm}
\stackunder{\hspace{-3.5cm}(d)}{\includegraphics[width=4.3cm]{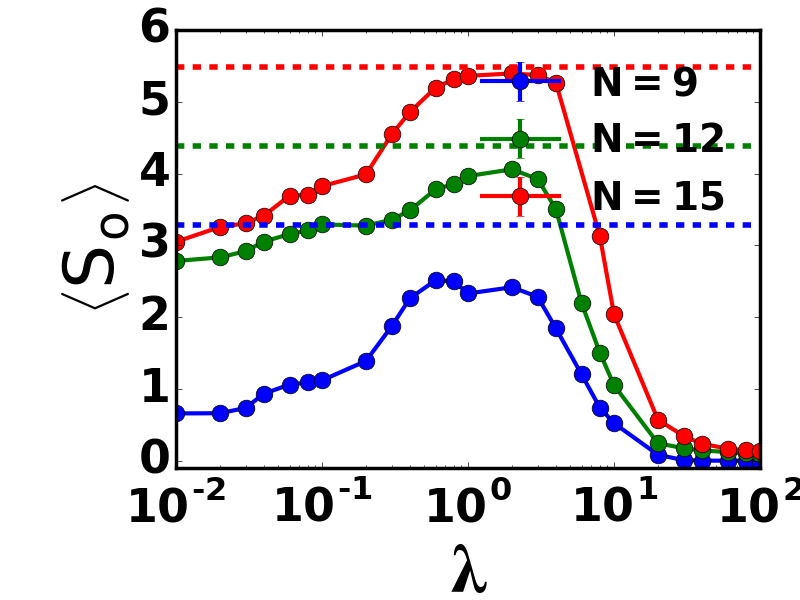}}
\caption{\label{fig18}~Occupation spectrum $\left\langle n_\alpha\right\rangle$ with scaled index $\alpha / N$ at fixed interaction strengths $V=1$ and disorder strengths (a) $\lambda=0.01$, (b) $\lambda=2$, and (c) $\lambda=100$, for different system sizes $N=9,12,15$ and fixed filling fraction $\nu=1 / 3$. (d) The average OPDM entropy $S_o$ with increasing strength of disorder $\lambda$. Dashed lines denote the maximal value of $S_o$. Averaging has been performed over the eigenstates in the energy window $\varepsilon=[0.54,0.57]$ and using $400,200$ and $50$ disorder realizations for system sizes $N=9,12$ and $15$, respectively.}
\end{figure}


We next study the eigenvector properties with the help of MIPR and the
OPDM. We study MIPR at a fixed interaction strength $V=1$ and various
disorder strengths $\lambda$ as shown in Fig.~\ref{fig16}. For a fixed
filling fraction $\nu=1 / 3$, we extract the exponent $\gamma$ by
averaging the MIPR over the states belonging to the energy window
$[\varepsilon-0.01, \varepsilon+0.01]$, where $\varepsilon=0.1,0.2,
\ldots, 0.9$. In the low disorder regime, $\lambda=0.01$ (see
Fig.~\ref{fig16}(a)), we observe a nonergodic mixed phase owing to
the spread of the exponent $\gamma$ over a wide range
$0.54<\gamma<0.80$. In the intermediate disorder case, $\gamma$ has a
significantly higher magnitude, which signifies thermal-like behaviour
(see Fig.~\ref{fig16}(b)). At high disorder strength $\lambda=100$,
the exponent $\gamma$ has a small magnitude $\approx 0.07$ over the
entire spectrum indicating an MBL-like phase, as shown in
Fig.~\ref{fig16}(c). In Fig.~\ref{fig17}, we have plotted the exponent
$\gamma$ at different disorder strengths and observe that in the low
disorder regime, $\gamma \approx 0.8$ corresponding to the energy
window about $\varepsilon=0.9$, which signifies the presence of
thermal-like states that contribute to the mixed nonergodic
behaviour. We observe that the intermediate and high disorder regimes
show thermal-like and MBL-like behaviour, respectively, over the
entire spectrum.

We next study the OPDM with the help of the occupation spectrum
$\left\langle n_\alpha\right\rangle$ at different disorder strengths
$\lambda=0.01, 2$ and $100$ (see Figs.~\ref{fig18}(a)--\ref{fig18}(c))
and over the energy window $\varepsilon=\left[0.54,0.57 \right]$. In
the intermediate regime (Fig.~\ref{fig18}(b)), we observe that with
increasing system size, $\left\langle n_\alpha\right\rangle$ spreads
about the filling fraction $\nu =1/3$ with a characteristic inverse
system-size variation on either side of the critical value of the
$\frac{\alpha}{N}$, indicating a thermal phase. In contrast, in the
high disorder regime (Fig.~\ref{fig18}(c)), $\left\langle
n_\alpha\right\rangle$ reaches close to $0$ and $1$, indicating
localized behaviour of the single-particle states. Hence, the phase is
MBL-like. In the low disorder regime ((Fig.~\ref{fig18}(a))), the
occupation spectrum is neither spread about the thermal value
$\left\langle n_\alpha\right\rangle = \nu =1/3$ nor does it reach
close to $0$ and $1$ (like in MBL), thus indicating mixed nonergodic
behaviour. The OPDM entropy $S_o$ (see in Fig.~\ref{fig18}(d))
signifies a nonergodic phase in the low disorder regime as it neither
reaches the thermal value (dashed lines) nor the MBL value ($0$). In
the intermediate disorder case, $S_o$ reaches its thermal value
denoted by dashed lines, especially for $N=15$, while it approaches
$0$ in the high disorder case, indicating MBL-like behaviour.

\begin{figure}
\centering \stackunder{}{\includegraphics[
    width=0.5\textwidth]{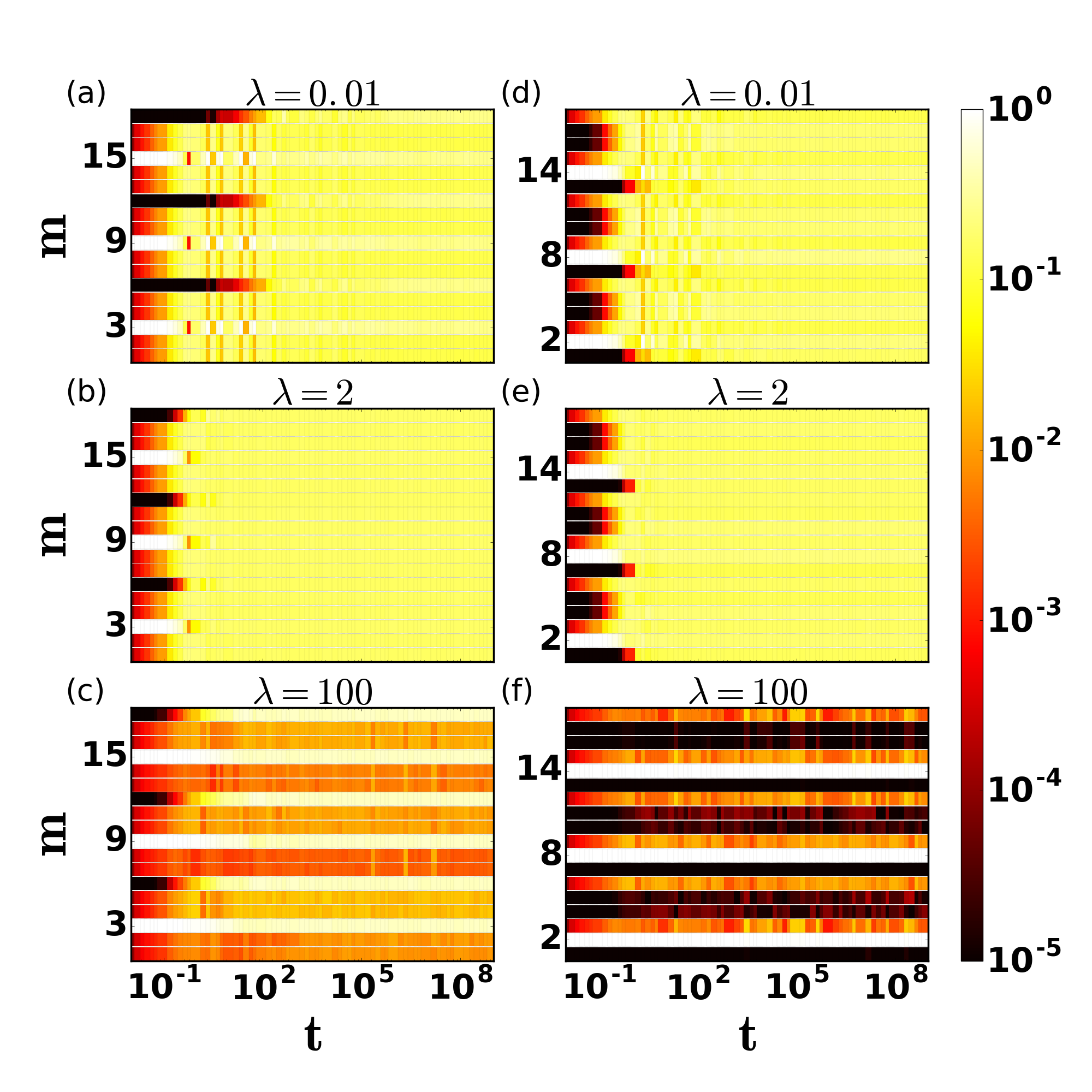}}
\caption{\label{fig19}~In the antisymmetric case, particle density (whose value is represented by a colour according to the code shown) as a function of time $t$, where $m$ is the site index, for the initial state given by Eq.~\ref{i1} for interaction strength $V=1$ and increasing disorder strengths (a)~$\lambda=0.01$, (b)~$\lambda=2$, (c)~$\lambda=100$. Corresponding plots (d)--(f)~show the evolution of the particle density for the initial state given by Eq.~\ref{i2}. $N=18$, $\nu=1/6$ and $100$ disorder realizations have been considered for all cases. }
\end{figure}

\begin{figure}
\centering
\stackunder{\hspace{-3.0cm}(a)}{\includegraphics[width=4.2cm]{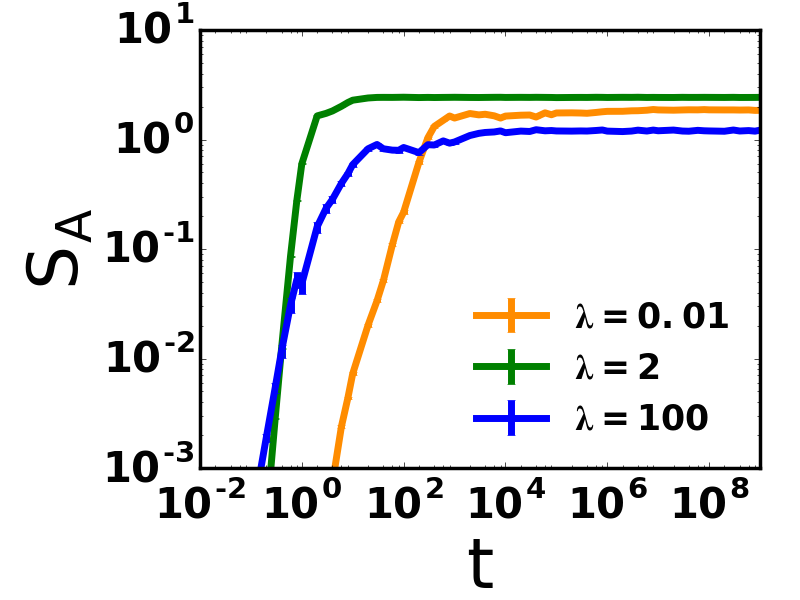}}
\stackunder{\hspace{-3.0cm}(b)}{\includegraphics[width=4.2cm]{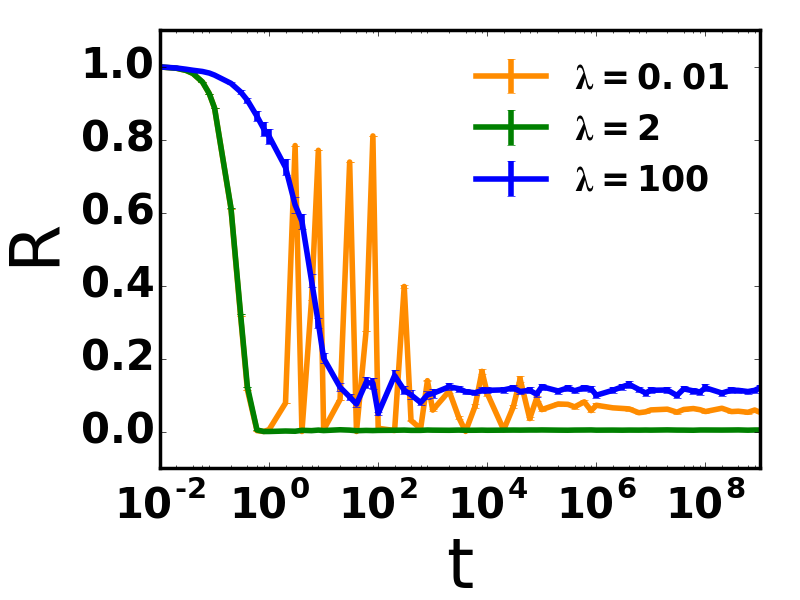}}
\vspace{-0.4cm}

\stackunder{\hspace{-3.0cm}(c)}{\includegraphics[width=4.2cm]{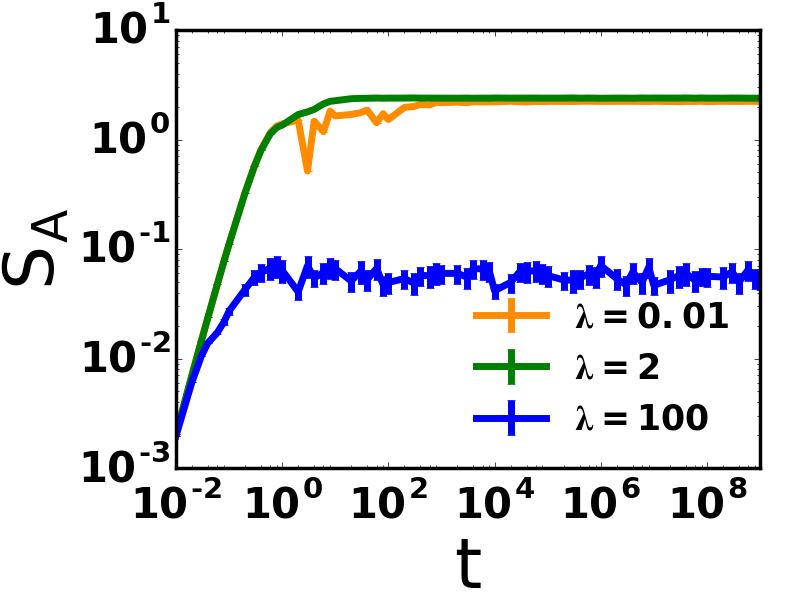}}
\stackunder{\hspace{-3.0cm}(d)}{\includegraphics[width=4.2cm]{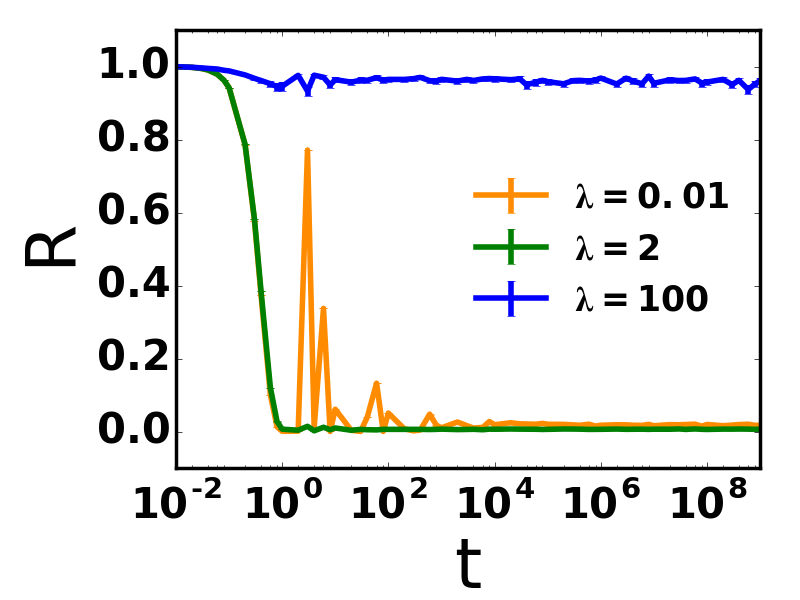}}
\caption{\label{fig20}~In the antisymmetric case, (a)~the entanglement
  entropy $S_A$ for a subsystem of size $N_A=N / 3$ and (b)~return
  probability $R$ as a function of time $t$ for the initial state
  given by Eq.~\ref{i1}. (c)~Entanglement entropy $S_A$ for a
  subsystem of size $N_A=N / 3$ and (d)~return probability $R$ as a
  function of time $t$ for the initial state given by
  Eq.~\ref{i2}. The interaction strength is fixed as $V=1$ for various disorder strengths $\lambda$. For all plots, $N=18, \nu=1
  / 6$, and the number of disorder realizations is $100$.}
\end{figure}

\begin{figure}[b]
\centering
\stackunder{\hspace{-3.8cm}(a)}{\includegraphics[ width=4.2cm]{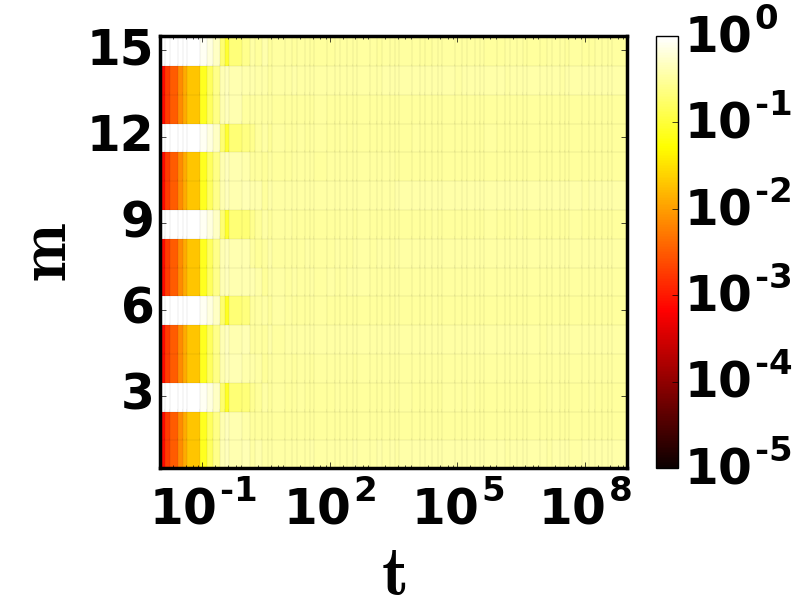}}
\stackunder{\hspace{-3.8cm}(b)}{\includegraphics[width=4.2cm]{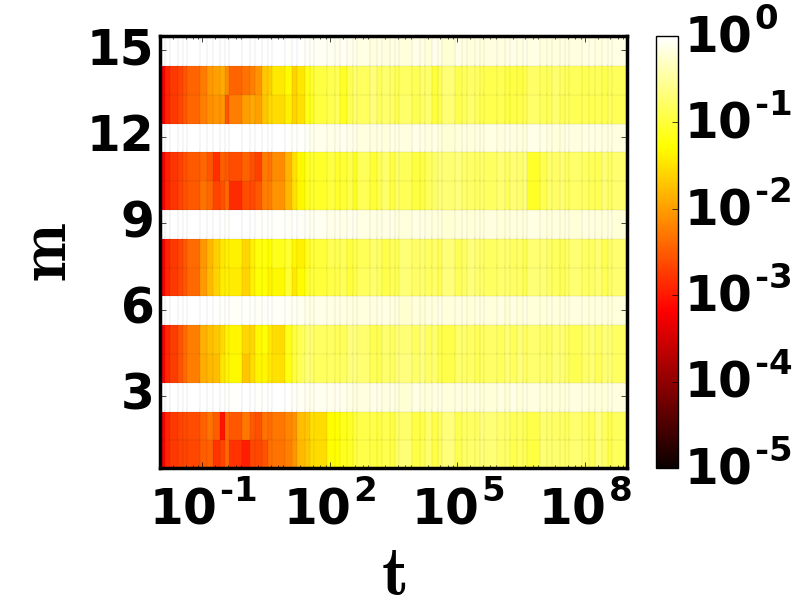}}
\vspace{-0.3cm}

\stackunder{\hspace{-3.8cm}(c)}{\includegraphics[width=4.2cm]{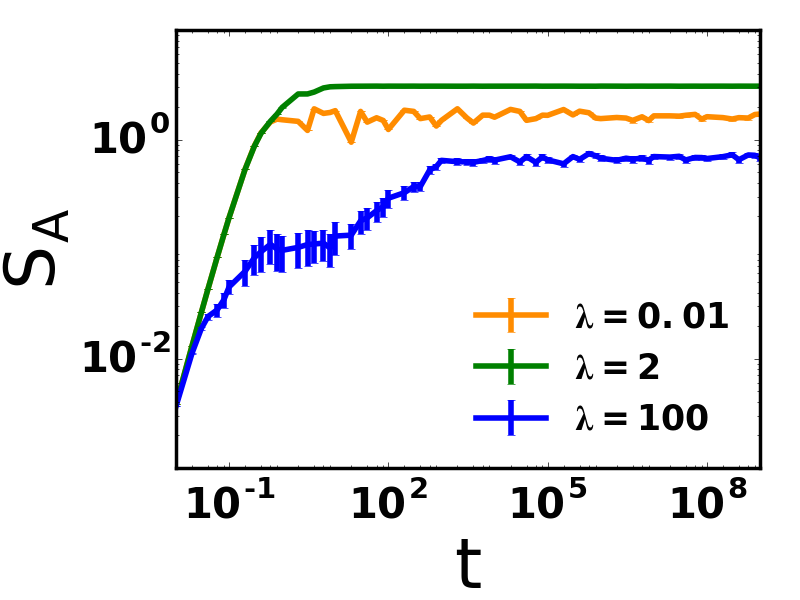}}
\stackunder{\hspace{-3.8cm}(d)}{\includegraphics[width=4.2cm]{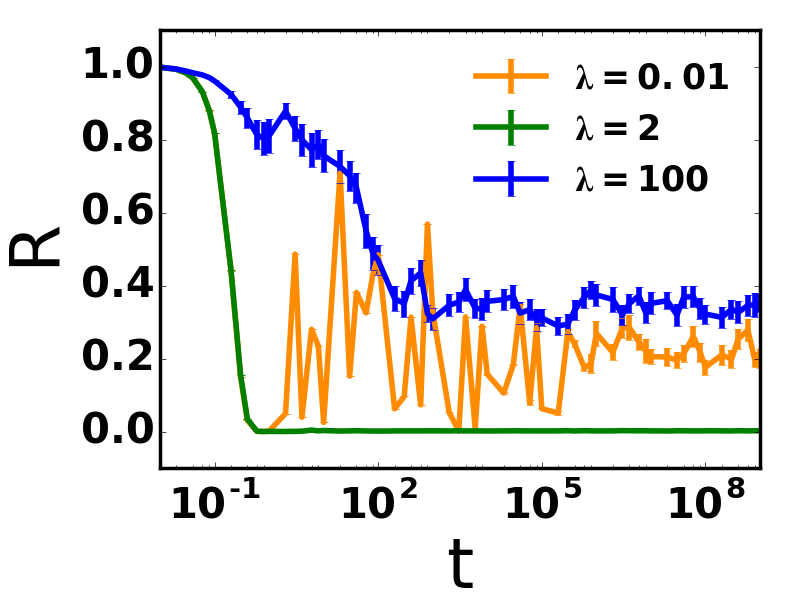}}

\vspace{-0.5cm}
\stackunder{\hspace{-3.8cm}(e)}{\includegraphics[width=4.2cm]{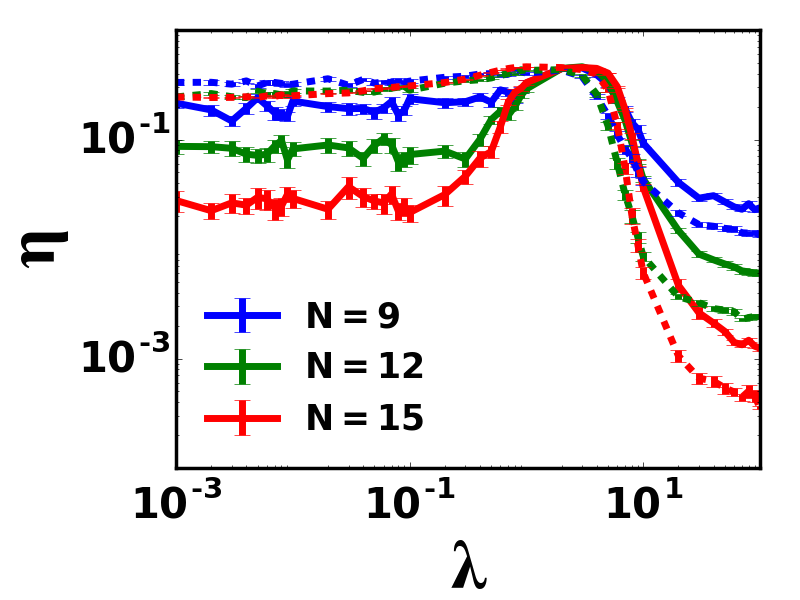}}
\stackunder{\hspace{-3.8cm}(f)}{\includegraphics[width=4.2cm]{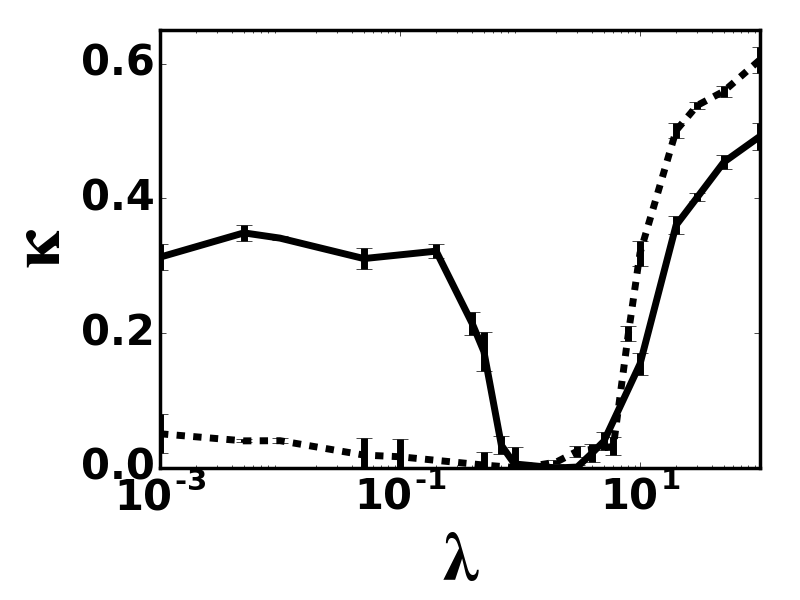}}
\caption{\label{fig21}~The particle density (whose value is represented
  by a colour according to the code shown) as a function of time $t$,
  where $m$ is the site index for the initial state given by
  Eq.~\ref{i3} for disorder strengths (a)~$\lambda=2$ and
  (b)~$\lambda=100$. (c)~Entanglement entropy $S_A$ for a subsystem of
  size $N_A=N / 3$ and (d)~return probability $R$ as a function of
  time $t$. Here the interaction strength is fixed to $V=1$, and the
  system size is $N= 15$ with filling fraction $\nu=1 / 3$, and
  averaging has been done over $50$ disorder realizations. (e)~In the
  long time limit $t=10^9$, NPR $\eta$ as a function of disorder
  strength $\lambda$ for various system sizes $N$, solid lines
  correspond to initial state given by Eq.~\ref{i3} while dashed lines
  correspond to initial state given by Eq.~\ref{i4}. (f)~The scaling
  exponent $\kappa$ as a function of disorder strength $\lambda$. Here $V=1$ and $\nu=$
  $1 / 3$, and the number of disorder realizations is at least $50$
  for all the system sizes.}
\end{figure}

\subsection{Nonequilibrium dynamics}
We next study many-body non-equilibrium dynamics with several measures
such as particle density, entanglement entropy and return
probability. We first consider the initial state given by
Eq.~\ref{i1}, a product state for a system size $N=18$ and filling
fraction $\nu=1/6$, with particles occupying the $c-$sites of
alternate unit cells. From the evolution of the particle density shown
in Fig.~\ref{fig19}(a)--\ref{fig19}(c), we observe that for all
disorder strengths, the CLS-like behaviour persists at early times
($t<1$). However, for time $t > 1$, while the particle density spreads
uniformly over the entire lattice in the low and intermediate disorder
regimes, it shows a comparatively less ergodic behaviour in the high
disorder case. We also study the dynamics for the second initial state
given by Eq.~\ref{i2}. For low and intermediate disorder strengths
(see Fig.~\ref{fig19}(d)--\ref{fig19}(e)), we observe that the
particle density is uniformly spread over all the sites indicating
ergodic behaviour. In contrast, at higher disorder strength $\lambda =
100$ (see Fig.~\ref{fig19}(f)), the particle density is significantly
localized over the initially occupied sites.

The entanglement entropy and return probability dynamics for both the
initial configurations are shown in Fig.~\ref{fig20}. In the case of
the initial state corresponding to Eq.~\ref{i1}, $S_A$ has near-zero
magnitude at early times, as shown in Fig.~\ref{fig20}(a). However,
the low and intermediate disorder regimes saturate to a large
magnitude after the initial transient. At higher disorder strengths
$(\lambda=100)$, we observe that the behaviour is relatively more
localized with the entanglement entropy $S_A$ (see
Fig.~\ref{fig20}(a)) saturating to a lower magnitude. From the return
probability dynamics as shown in Fig.~\ref{fig20}(b), while for low
and high disorder $R(t)$ saturates to a finite value indicating
nonergodic behaviour, in the intermediate disorder regime, $R(t)
\approx 0$ which signifies a thermal-like phase. For the second
initial state given by Eq.~\ref{i2}, after the initial transient, in
the low disorder regime, we observe oscillatory behaviour followed by
a subdiffusive growth after which $S_A$ saturates near the thermal
value(see Fig.~\ref{fig20}(c)). The behaviour of $S_A$ in the
intermediate disorder regime is similar to the low disorder case,
except that the oscillatory part is absent. After the transient, there
is a sub-diffusive increment in $S_A$, which saturates to a large
value indicating delocalization in the many-body system. In contrast,
in the high disorder region, it saturates to a low value signifying
MBL. We also study the evolution of the return probability, as shown in
Fig.~\ref{fig20}(d). In the long time limit, it saturates to $0$ in
the low and intermediate disorder regimes indicating thermal-like
behaviour. At the same time, it is close to $1$ in the high disorder
regime, which is a signature of many-body localization.

We next study the non-equilibrium dynamics of the system for the two
initial states by considering filling fraction $\nu=1/3$. The
evolution of the particle density is shown in
Fig.~\ref{fig21}(a)--~\ref{fig21}(b), at the interaction strength
$V=1$ and for disorder strengths $\lambda=2$ and $100$. We observe
mixed nonergodic behaviour at low disorder strength (not shown
here). In contrast, we see ergodic behaviour at intermediate disorder
strengths, with the particle density spread uniformly over all the
sites. At high disorder strength $\lambda=100$, the particle density
is significantly localized over the initially occupied sites. We also
study the dynamics of entanglement entropy $S_A$ as shown in
Fig.~\ref{fig21}(c) at $V=1$. After the initial transient, $S_A$ shows
a subdiffusive growth in both the low and intermediate disorder
regimes; however, it saturates to a large value for $\lambda=2$,
indicating an ergodic phase, while for $\lambda=0.01$, it saturates to
a comparatively lower value indicating weak delocalization in the
system. In the high disorder case, after the initial transient, $S_A$
saturates to a sub-thermal value indicating localization. From the
dynamics of the return probability (see Fig.~\ref{fig21}(d)), we
observe that it saturates to a finite value in the low and high
disorder phases. In contrast, it saturates close to $0$ in the
intermediate phase indicating thermal behaviour.

We next study the NPR~\cite{PhysRevB.87.134202} in the long time limit
for the initial states given by Eq.~\ref{i3} (solid lines) and
Eq.~\ref{i4} (dashed lines) as shown in Fig.~\ref{fig21}(e). In the
case of the initial state given by Eq.~\ref{i3}, while $\eta$ is
system size dependent in the low and high disorder regimes, it shows
system size independence in the intermediate disorder case. We then
extract the exponent $\kappa$ and plot it as a function of disorder
strength $\lambda$ in Fig.~\ref{fig21}(f). In the low disorder case,
$\kappa$ lies between $0$ and $0.5$, indicating nonergodic behaviour
while in the intermediate regime $\kappa=0$, indicating a thermal
phase. In contrast, in the high disorder regime, we observe $\kappa
\approx 0.5$, which is a signature of many-body localization. For the
other initial state given by Eq.~\ref{i4}, in the low disorder regime,
$\eta$ tends to be system size-independent (see Fig.~\ref{fig21}(e)),
with $\kappa$ close to $0$ (Fig.~\ref{fig21}(f)), which indicates that
it is close to the ergodic phase. In the intermediate regime, the
phase is thermal as $\eta$ is system size independent and
$\kappa=0$. In the higher disorder regime, $\eta$ is system size
dependent with $\kappa$ close to $0.5$, indicating an MBL-like phase.

\section{Conclusion}\label{sec:level7}
In this work, we have systematically investigated the single-particle
dynamics and the interplay of many-body interactions and
quasiperiodic AA disorder in the one-dimensional ABF diamond
lattice. We find that the compact localized states observed for the
clean system and when the disorder is applied symmetrically in the
single particle case~\cite{PhysRevB.106.205119} sustain quantum caging
even in a non-equilibrium dynamical set-up. In contrast, in the long
time limit, the wave function spreads over the entire lattice in the
presence of antisymmetric disorder. This can be attributed to the loss
of compact localization and the presence of multifractal eigenstates
in the static case~\cite{PhysRevB.106.205119}.

In the presence of interactions and zero disorder in the system,
nonergodic phases are observed at all interaction strengths. In
general, non-equilibrium dynamics support the findings from the static
case. However, the many-body system manifests quantum caging for
specially engineered initial states. When our interacting many-body
system is subjected to symmetric disorder, nonergodic phases are
observed at low and intermediate disorder strengths. In contrast, an
MBL-like phase is observed at higher disorder strengths. Studying
non-equilibrium dynamics, we find nonergodic regimes in the case of
low and intermediate disorder strengths, while localization
characteristics dominate in the high disorder case. Quantum caging
behaviour is supported for specific initial configurations,
independent of the strength of interaction or disorder.

The antisymmetric application of disorder in the presence of
interactions in the system result in three distinct phases: a
nonergodic mixed phase at low disorder strengths, a thermal phase at
intermediate disorder strengths and an MBL-like phase at higher
strengths of disorder. Even in the mixed nonergodic phase, some
states show thermal-like behaviour. A study of the non-equilibrium
dynamics shows that for different initial states, in the low disorder
regime, a mixed phase exists with a varying magnitude of
non-ergodicity; for the intermediate disorder, the phase is always
thermal. In the high disorder case, the phase shows varying magnitudes
of non-ergodicity, inclined towards many-body localization.

We also want to remark on the case when the on-site disorder is chosen from a uniform uncorrelated random distribution in the presence of interactions. Interestingly, for both the symmetric and antisymmetric cases, the resulting phase is similar in the low, intermediate and high disorder strengths to those obtained when applying the quasiperiodic disorder. We conclude that the observed many-body phases are due to the symmetry of the applied disorder as reported previously in the single particle case~\cite{PhysRevB.106.205119}. Also, in the antisymmetric case, the presence of mixed non-ergodic, thermal and MBL-like phases at the low, intermediate and high disorder strengths, respectively, is qualitatively similar to phases that emerge as a result of the application of the uniform disorder on all sites as reported in~\cite{PhysRevResearch.2.043395}.  
 
Thus, our work shows that the interplay of quasiperiodic disorder,
interactions and flat-band structure in the diamond lattice results in
an exciting phase diagram. Exploring distinctive phases in other
interacting and disordered flat band systems would be an interesting
direction for further research. With the recent surge in the
experimental study of engineered flat-band systems, such phases could
be realized in optical lattices for cold atoms.


\begin{acknowledgments}
We are grateful to Ajith Ramachandran for the careful reading of the
manuscript and discussions. We are grateful to the High-Performance
Computing (HPC) facility at IISER Bhopal, where large-scale
computations of this project were run. A.A. is grateful to the Council
of Scientific and Industrial Research (CSIR), India, for her PhD
fellowship. N.R. acknowledges support from the Indian Institute of
Science under the IoE-IISc fellowship program. A.S. acknowledges
financial support from SERB via the grant (File Number:
CRG/2019/003447) and DST via the DST-INSPIRE Faculty Award
[DST/INSPIRE/04/2014/002461].
\end{acknowledgments}

\section*{APPENDIX: ENTANGLEMENT ENTROPY}
In this section, we analyze the effect of interactions on the ABF
diamond lattice by calculating the half-chain entanglement entropy
$S_A$ of all the many-body eigenstates. We first discuss the case when
interactions are turned on in the disorder-free model, and then
consider cases where disorder is applied in a symmetric and
antisymmetric manner in the presence of interactions.

\begin{figure}[b]
\centering
\stackunder{}{\includegraphics[ width=5.5cm]{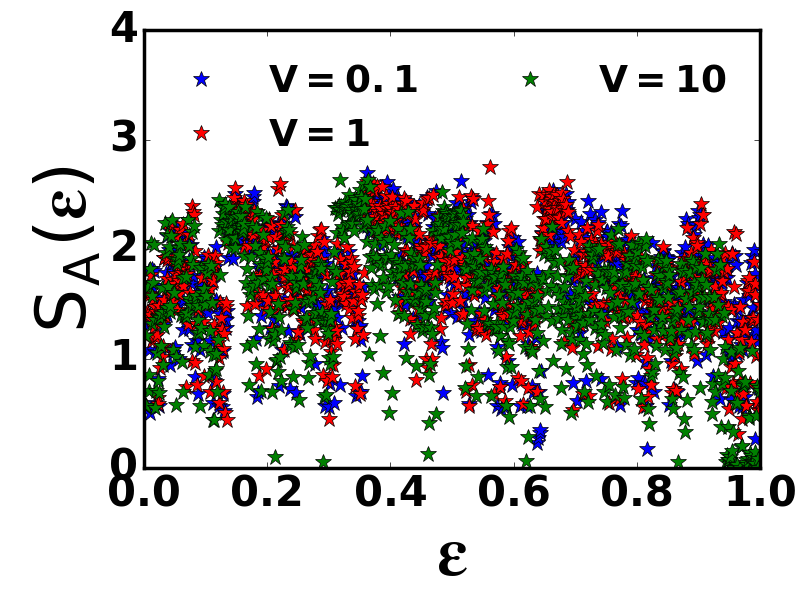}}
\caption{\label{fig22} The half-chain entanglement entropy $S_A(\epsilon)$ of all the eigenstates as a function of the fractional eigenstate index $\epsilon$ for $N=18$ with $\nu=1/6$ for interaction strengths $V=0.1, 1$ and $10$.}
\end{figure}

Fig.~\ref{fig22} shows the half-chain entanglement entropy $S_A$ of
all the many-body eigenstates at various interaction strengths $V$ for
a system size $N=18$ and filling fraction $\nu=1/6$. For all
interaction strengths, $V=0.1, 1$ and $10$, while a significant
fraction of the eigenstates shows a large magnitude of the
entanglement entropy, it does not vary smoothly with the fractional
eigenstate index $\epsilon$. This indicates the presence of a
nonergodic phase, which agrees with the MIPR and OPDM results shown
in Section~\ref{sec:level4}.

\begin{figure}
\centering
\stackunder{\hspace{-4cm}(a)}{\includegraphics[width=4.3cm]{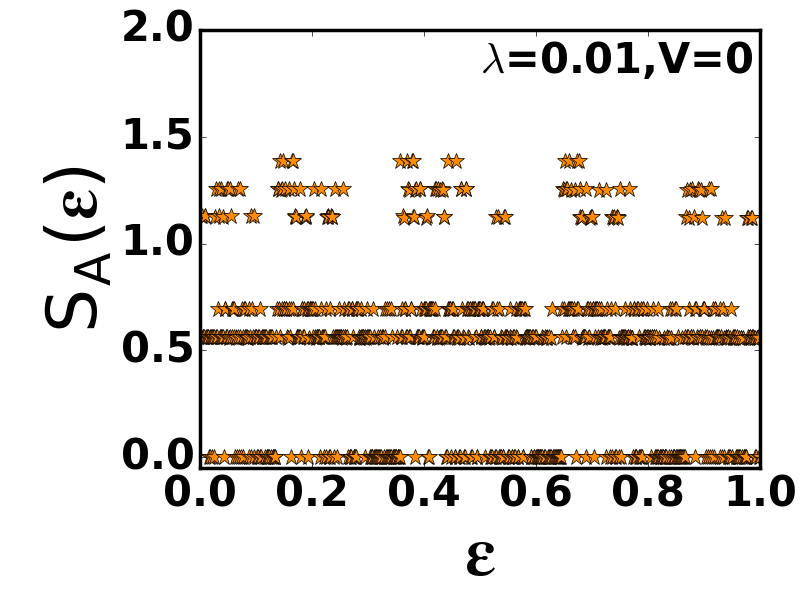}}\hspace{-1mm}
\stackunder{\hspace{-4cm}(b)}{\includegraphics[width=4.3cm]{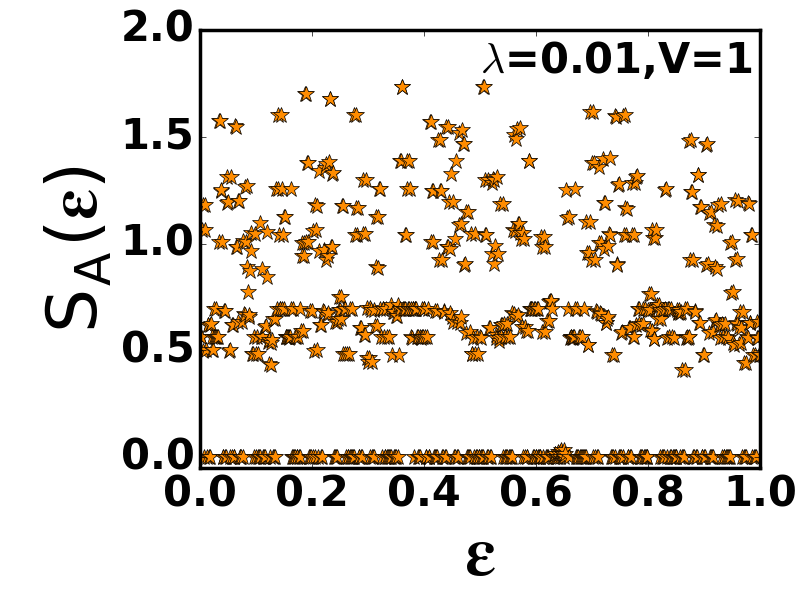}}
\vspace{-0.3cm}

\stackunder{\hspace{-4cm}(c)}{\includegraphics[width=4.3cm]{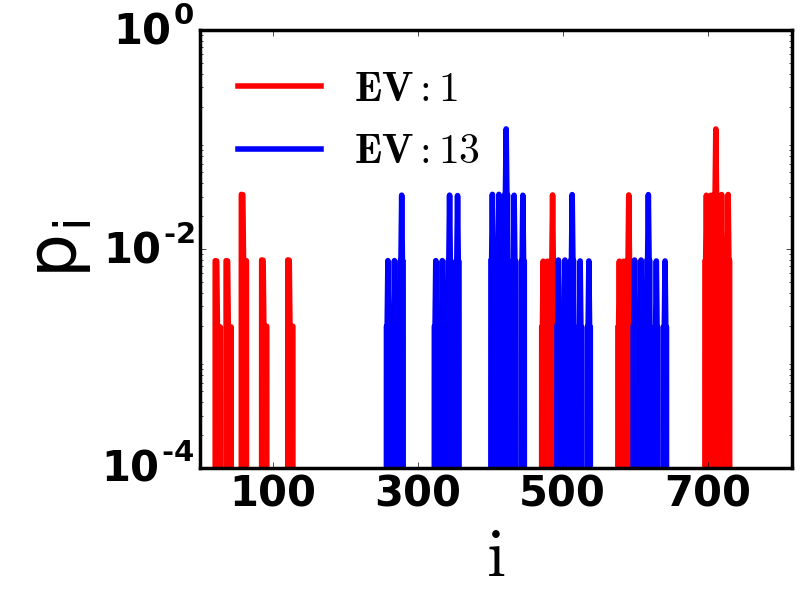}}\hspace{-1mm}
\stackunder{\hspace{-4cm}(d)}{\includegraphics[width=4.3cm]{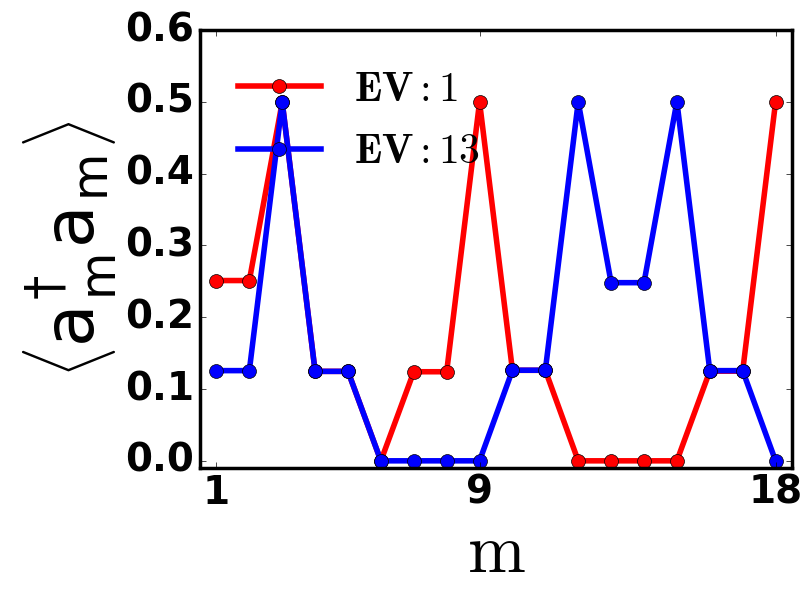}}
\caption{\label{fig23} The half chain entanglement entropy
  $S_A(\epsilon)$ for all the eigenstates as a function of the
  fractional eigenstate index $\epsilon$ for disorder strength
  $\lambda=0.01$ and interaction strengths (a)~$V=0$ and
  (b)~$V=1$. (c)~The wavefunction probability of two states indexed as
  $EV:1$ with $S_A \neq 0$ and $EV:13$ with $S_A = 0$. Here $i=1,\ldots,D$,
  where $D$ is the dimension of the particle number constraint Hilbert
  space. (d)~The corresponding particle density $\left\langle
  a_m^{\dagger} a_m \right\rangle$, where $m$ is site index at
  $V=0$. For all the cases, $N=18$, filling fraction $\nu=1/6$ and
  single disorder realization has been considered.}
\end{figure}


\begin{figure}[b]
\centering
\stackunder{}{\includegraphics[ width=5.5cm]{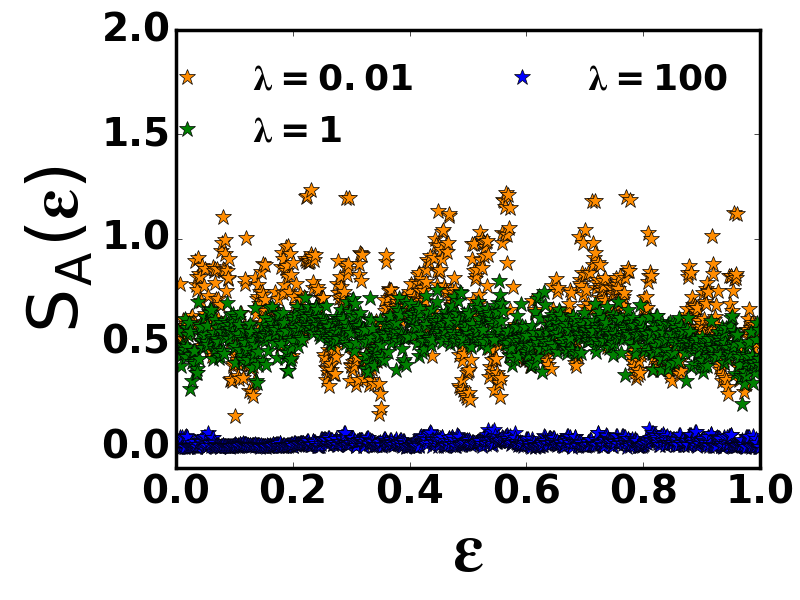}}
\caption{\label{fig24} In the symmetric case, half-chain entanglement entropy $S_A(\epsilon)$ for all the eigenstates as a function of the fractional eigenstate index $\epsilon$ for disorder strengths $\lambda=0.01, 1$ and $100$. The averaging has been performed over $50$ disorder realizations.}
\end{figure}

\begin{figure}
\centering
\stackunder{\hspace{-4cm}(a)}{\includegraphics[ width=4.34cm]{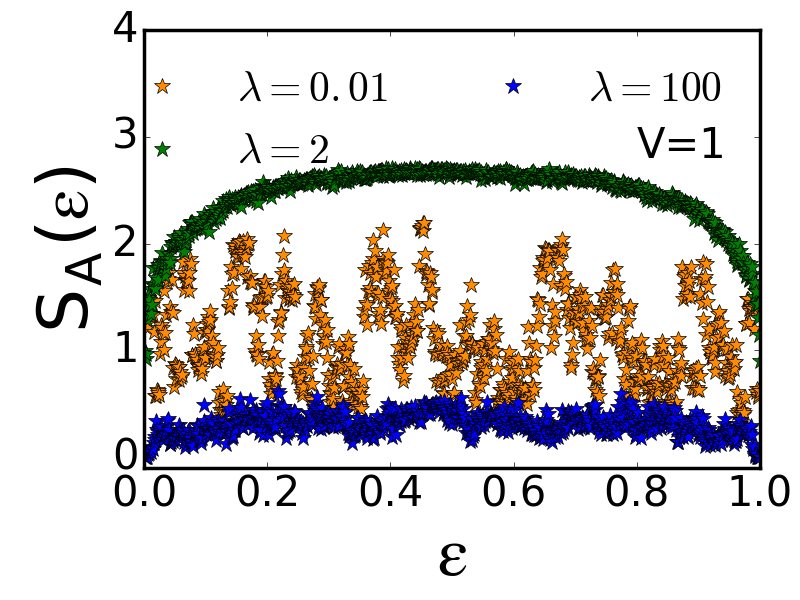}}\hspace{-2mm}
\stackunder{\hspace{-4cm}(b)}{\includegraphics[width=4.34cm]{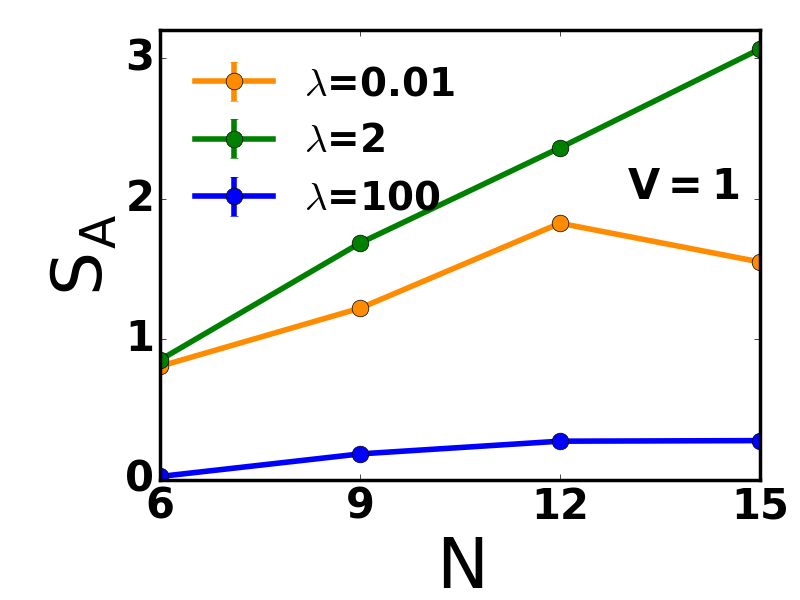}}
\vspace{-0.3cm}

\stackunder{\hspace{-4cm}(c)}{\includegraphics[width=4.34cm]{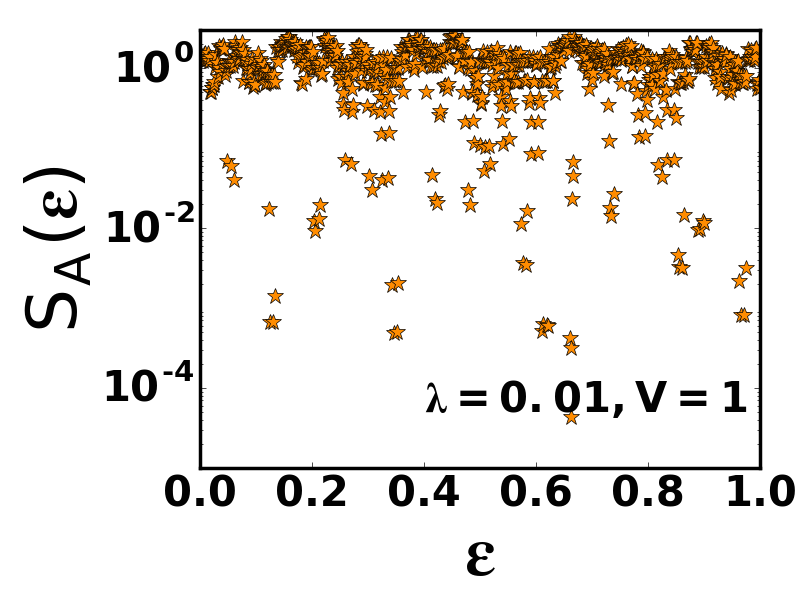}}\hspace{-2mm}
\stackunder{\hspace{-4cm}(d)}{\includegraphics[width=4.34cm]{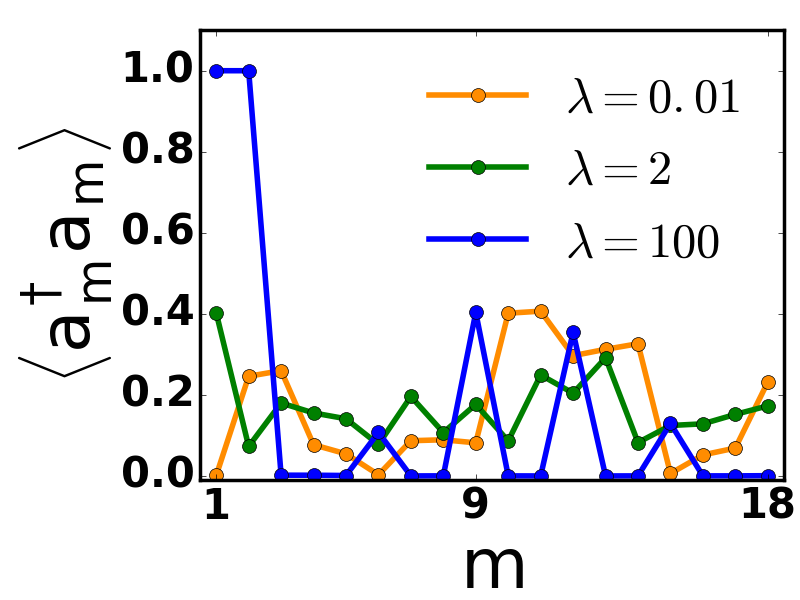}}
\caption{\label{fig25}(a)~The half-chain entanglement entropy $S_A(\epsilon)$ of all the eigenstates as a function of the fractional eigenstate index $\epsilon$ for $N=18$ with $\nu=1 / 6$ for various disorder strengths $\lambda$ and fixed interaction strength $V=1$. Here averaging has been performed over $100$ disorder realizations. (b)~Entanglement entropy $S_A$ averaged over the eigenstates in the energy window $\varepsilon=[0.54,0.57]$ with system size $N$, filling fraction $\nu=1 / 3$ and subsystem size $N_A=N / 3$. Here interaction strength $V=1$ and disorder realizations is $100$. (c)~The half-chain entanglement entropy $S_A(\epsilon)$ of all the eigenstates as a function of the fractional eigenstate index $\epsilon$ for $N=18$, filling fraction $\nu=1 / 6$, disorder strength $\lambda=0.01$, interaction strength $V=1$ and single disorder realization. (d)~Particle density $\left\langle a_m^{\dagger} a_m\right\rangle$ with site index $m$, filling fraction $\nu=1 / 3$ for the infinite temperature state at $\theta=0$ and $V=1$.}
\end{figure}

Next, we study the application of symmetric disorder; we have reported
the presence of compactly localized states~\cite{PhysRevB.106.205119}
in the single-particle case. Interestingly, in the case of
non-interacting fermions, from the half-chain entanglement entropy, we
observe that the eigenstates can be broadly divided into two
categories (see Fig.~\ref{fig23}(a)), one with $S_A \neq 0$ and the
other with $S_A=0$. We choose a state corresponding to $S_A\neq 0$
with index $EV=1$ and another corresponding to $S_A = 0$, with index
$EV=13$. In Fig.~\ref{fig23}(c), we have plotted both eigenstates in
the particle number-constrained space with the index $i$ running over
$1,2,\ldots, D$ and observe non-zero amplitude only on a finite number of sites
. For the given state, we calculate the particle density
$\left\langle a_j^{\dagger}a_j \right\rangle$ (see Eq.~\ref{eq9}) on
all the sites $j=1,2,\ldots,N$ as shown in Fig.~\ref{fig23}(d). In the
case of the state corresponding to $S_A = 0$, the particle density is
zero both at the subsystem boundary ($N/2$ or $N/2+1$) as well as the
system boundaries ($1$ or $N$) as periodic boundary conditions have
been considered thereby disassociating the two subsystems. However,
the same is not true for the states where the entanglement is
non-zero. This effect persists for all strengths of the disorder as
well as on the introduction of interactions (see
Fig.~\ref{fig23}(b)). We conclude that this is a many-body effect in a
disordered system hosting compactly localized states. Also, a blind
disorder averaging the entanglement entropy would wash out this
behaviour, as shown in Fig.~\ref{fig24}.

Fig.~\ref{fig25}(a) shows the half-chain entanglement entropy for all
the many-body eigenstates in the case of antisymmetric application of
disorder and interaction strength $V=1$. In the low disorder case,
$S_A$ shows both rises and dips, indicating the presence of a
nonergodic mixed phase. In the intermediate disorder regime, $S_A$
has a smooth dependence on the eigenstates with a large magnitude
indicating thermal behaviour. In contrast, for the higher disorder
$\lambda=100$, we observe $S_A$ with a meagre value indicating
MBL-like behaviour. These results agree with those discussed in
Section~\ref{sec:level6} from the study of MIPR and OPDM. In
Fig.~\ref{fig25}(b), we plot the entanglement entropy $S_A$ for a
fixed filling fraction $\nu=1/3$ and subsystem size $N_A=1/3$ with
increasing system size $N$ in the energy window $\varepsilon
=[0.54,0.57]$~\cite{PhysRevB.91.081103}. We find that the intermediate
disorder case ($\lambda=2$) follows a volume law scaling while an area
law-like behaviour is seen in the high disorder regime
($\lambda=100$). In the low disorder regime, $S_A$ initially increases
and eventually saturates as a function of system size, thus indicating
nonergodic behaviour. We also study a single disorder realization of
the half-chain entanglement entropy with $\lambda=0.01$ and $V = 1$
(see Fig.~\ref{fig25}(c)). Unlike the symmetric
disorder case, we observe that $S_A \neq 0$ for any eigenstate. We also study the
particle density (see Fig.~\ref{fig25}(d)) for the infinite
temperature state and observe that in the low disorder case
($\lambda=0.01$), it is unevenly spread over the lattice sites
indicating nonergodic behaviour. In contrast, at $\lambda=2$, it
spreads out uniformly, showing thermal behaviour, and at higher
disorder strengths $\lambda=100$, it is localized over a few sites.

\bibliography{dia_int}

\end{document}